\documentstyle[12pt]{article}
\newcommand{\xslash}{\not \! x}
\newcommand{\kslash}{\not \! k}
\newcommand{\pslash}{\not \! p}
\newcommand{\delslash}{\not \! \partial}

\begin{document}

\begin{flushright}{UT-03-07 \\ YITP-SB-03-11}
\end{flushright}
\vskip 0.5 truecm

\begin{center}
{\Large{\bf Topological anomalies from the path integral measure 
in superspace}}
\end{center}
\vskip .5 truecm
\centerline{\bf Kazuo Fujikawa\footnote{
         fujikawa@phys.s.u-tokyo.ac.jp} }
\vskip .4 truecm
\centerline {\it Department of Physics,University of Tokyo}
\centerline {\it Bunkyo-ku,Tokyo 113,Japan}
\vskip 0.5 truecm
\centerline{\bf Peter van Nieuwenhuizen\footnote{
 vannieu@insti.physics.sunysb.edu}} 
\vskip .4 truecm 
\centerline{\it C.N. Yang Institute for Theoretical Physics} 
\centerline{\it State University of New York, Stony Brook, NY 
11794-3840, USA} 
\vskip 0.5 truecm

\makeatletter
\@addtoreset{equation}{section}
\def\theequation{\thesection.\arabic{equation}}
\makeatother

\setcounter{footnote}{0}

\begin{abstract}
A fully quantum version of the Witten-Olive analysis of the 
central charge in the $N=1$ Wess-Zumino model in $d=2$ with a 
kink solution is presented by using path integrals in superspace.
We regulate the Jacobians with heat kernels in superspace, and 
obtain all superconformal anomalies as one Jacobian factor.
The conserved quantum currents differ from the Noether currents 
by terms proportional to field equations, and these terms 
contribute to the anomalies. 
We identify the particular variation of the superfield which 
produces the central charge current and its  
anomaly; it is the variation of the auxiliary field.
The quantum supersymmetry algebra which includes the 
contributions of superconformal anomalies is derived by using 
the  Bjorken-Johnson-Low method instead of 
semi-classical Dirac brackets. We confirm earlier results that 
the BPS bound remains saturated at the quantum level due to 
equal anomalies in the  energy and central charge. 
\end{abstract}

%\large

\section{Introduction and brief summary}

Supersymmetry and topology are intimately linked. 
For example, instantons play an important role in the effective
action for rigidly supersymmetric (susy) models\cite{R1}.
The Donaldson invariants, which characterize topological 
properties of compact manifolds, can be computed by using a
particular topological field theory which is obtained by 
twisting a
Euclidean supersymmetric $N=(2,2)$ model\cite{R2}. We shall 
consider here the surface terms  in the supersymmetry
 algebra\cite{R3} which  form the central charges. 

The supersymmetry algebra of the kink, an $N=(1,1)$ rigidly 
supersymmetric  model in $1+1$ dimensions with a soliton
 solution,
reads as follows at the classical level~\cite{R3}
\begin{equation}
\{Q_{cl},Q_{cl} \}=2H_{cl}-2Z_{cl}
\end{equation}
Here $Q_{cl}$ is the classical supersymmetry charge which leaves
the classical kink solution invariant and,  properly extended
to the quantum level, should leave the kink vacuum invariant.
$H_{cl}$ is the classical Hamiltonian which gives the 
classical mass of the kink solution $\varphi_{K}(x)$, and 
$Z_{cl}$ is the integral of a total
derivative 
\begin{equation}
Z_{cl}=\int^{\infty}_{\infty}U(\varphi_{K})\partial_{x}
\varphi_{K} dx
\end{equation}
which is non-vanishing because the kink solution has a 
topological twist ($\varphi_{K}(\infty)$ differs from 
$\varphi_{K}(-\infty)$ ). The result in (1.1) can be derived 
by using semi-classical Dirac brackets.

In the 1970's and 1980's solitons were studied in 
detail~\cite{Dashen:1974cj},
and the issue whether for supersymmetric solitons $Z$ is modified by quantum 
corrections was studied in several articles, with conflicting 
results~\cite{kaul}. The kink model breaks conformal symmetry 
explicitly and is nonintegrable, and hence methods used for
exactly soluble models were of no avail.
Six years ago the issue whether the BPS bound 
$H_{cl}=Z_{cl}$ remains satisfied at the 
quantum level was again raised~\cite{Rebhan:1997iv},
and subsequently in a series of articles by several authors the 
quantum
corrections to $\langle H\rangle$ and $\langle Z\rangle$ were 
 calculated, where by $\langle H\rangle$ we mean the
expectation value of the quantum Hamiltonian in the kink 
vacuum, and similarly for $\langle Z\rangle$. It was found that
even though $Z$ is classically the integral of a total 
divergence, there are 
nonvanishing quantum corrections to $\langle Z\rangle$ which are equal 
to those to $\langle H\rangle$, so that the BPS bound remains 
saturated at the quantum 
level~\cite{Nastase:1998sy, Shifman:1998zy, Graham:1998qq,
Rebhan:2002yw}. 
The nonvanishing corrections to $\langle Z\rangle$ come from
a new anomaly, whose existence was  
 conjectured in~\cite{Nastase:1998sy} and
subsequently found and evaluated in~\cite{Shifman:1998zy}.
This result was in fact in conflict with the result
of \cite{Graham:1998qq} where BPS saturation and nonvanishing 
quantum corrections to $\langle H\rangle$ and $\langle Z\rangle$
 were obtained apparently without the need for the anomalous 
term in $Z$ found in \cite{Shifman:1998zy}. However, as has been
 clarified recently in \cite{Rebhan:2002yw}, this was due to 
manipulations of unregularized expressions; consistent 
dimensional regularization indeed reproduces the anomaly in $Z$.
 The existence of an anomaly in $Z$ is therefore by now beyond 
doubt.
Qualitatively, the reason is that $Z$ is 
a composite 
operator which should be regularized at the quantum level by,
for example, point-splitting, and although both the nonanomalous
and the anomalous corrections to  the central charge 
density $\zeta^{0}(x)$ are still total divergences in this 
regularization scheme, 
the space integral of the anomalous contributions no longer 
vanishes, being
proportional to $\int_{-\infty}^{\infty}
U^{\prime\prime}(\varphi_{K})\partial_{x}\varphi_{K}dx $.
However, the details are quite subtle;
different regularization schemes give unexpected contributions 
to $\zeta^{\mu}$, which consist of non-anomalous and anomalous 
contributions. For example,
using ordinary ('t Hooft-Veltman) dimensional regularization,
parity violation due to massless chiral domain wall fermions in 
the extra dimension is responsible for the anomaly in the central
charge in 2 dimensions, 
whereas using dimensional reduction, 
the anomalies reside not in loop graphs but in the evanescent
counter terms which renormalize the 
currents~\cite{Rebhan:2002yw}. Using 
the higher space-derivative regularization scheme,
there is an extra term in the central charge current which
produces the anomaly~\cite{Shifman:1998zy}, while using heat 
kernel methods,
subtle boundary contributions produce 
anomalies~\cite{Bordag:2002dg}.

The discovery that an anomaly is present in the central charge
 has led to precise 
calculations which restore the BPS bound at the quantum level, 
but raises profound questions concerning  topological
symmetries at the quantum level. In~\cite{Rebhan:2002yw}, a 
preliminary analysis was made of ordinary and conformal 
multiplets of 
currents, and ordinary and conformal multiplets of anomalies; in 
particular, a conformal central charge current was identified 
whose divergence contained, in addition to terms due to 
explicit symmetry breaking, the anomaly in the central charge.
Thus the central charge contains an anomaly and is itself the 
anomaly of another current.

In this article, we intend to study the anomaly structure of the 
currents of the supersymmetric kink model in superspace.
We use the path integral formulation of anomalous Ward
identities~\cite{fuji1,fuji2}, and the present work
extends a superspace analysis of conformal anomalies in 
QCD in $3+1$ dimensions~\cite{konishi}\cite{shizuya}.
A superspace approach to the anomalies in the central charge 
and the kink energy was first given in~\cite{Shifman:1998zy}.
In that article several regularization schemes were used to
evaluate the one-loop corrections, in particular a higher 
derivative regularization scheme in superspace. We shall start
with a path integral approach, and compute the Jacobians 
for generalized supersymmetry transformations using a 
superspace heat kernel. This will lead to a multiplet of 
anomalies which contains the trace anomaly and the central 
charge anomaly in addition to the supersymmetry anomaly and 
other terms. 
Crucial in this approach is that careful regularization
of terms proportional to the field equation of the auxiliary
field yields nonvanishing contributions to the Ward identities.
In the literature, some articles deal directly with 
the central 
charge current while in some other articles the central charge 
 anomaly is obtained from a supersymmetry transformation of 
the conformal anomaly in the supercurrent.
 We shall derive 
the central charge anomaly both directly by evaluating the 
Jacobian, and by 
a supersymmetry transformation of the conformal  anomaly
in the supercurrent, and show that the results agree with 
each other.

We begin with a local ($x$ {\bf and} $\theta$ dependent) 
supersymmetry transformation of the scalar superfield 
$\phi(x,\theta)$ and apply the Noether theorem in superspace.
 Using the 
quadratic part of the superspace action as regulator,
we obtain  a Ward identity in superspace (corresponding to 
 a hierarchy of
Ward identities in $x$-space) which contains the 
one-loop anomalies 
in certain quantum currents $\tilde{J}_{\mu}(x)$, 
$\tilde{T}_{\mu\nu}(x)$, $\tilde{\zeta}_{\mu}(x)$, in addition to
 explicit symmetry breaking terms. 
The use of a  
superfield formulation of heat kernels in strictly $d=2$ 
Minkowski 
space-time to regularize the Jacobians in the path integral 
formulation manifestly preserves ordinary rigid 
supersymmetry at all stages.
A subtle point 
is the proper identification of the quantum currents. The naive 
Noether currents 
are not conserved, but the Ward identities contain the 
currents $\tilde{J}_{\mu}(x)$, 
$\tilde{T}_{\mu\nu}(x)$, $\tilde{\zeta}(x)_{\mu}$ which are 
conserved,
 so these can be used to construct time-independent charges.
If we were to substitute the field equations in these conserved 
quantum currents,
we would obtain the Noether currents, but this is not allowed 
at the quantum level. In fact, the contributions proportional
to field equations produce  anomalies, as we already mentioned.

We also derive the supersymmetry algebra at the quantum level.
This requires a full quantum operator approach which incorporates
anomalies, rather than a semiclassical approach based on Dirac
brackets. Fortunately there exists such an approach:
 the Bjorken-Johnson-Low (BJL) method which automatically 
incorporates the effects of superconformal anomalies. 
The BJL method has been widely used for current 
algebras in the 1960's~\cite{bjorken}. 
It allows one to rewrite results obtained from path integrals
into operator relations. 
To go from the path integral results to operator results one 
must use field equations, but these field equations
sometimes  yield anomalies. The BJL method takes such anomalies 
into account, and this is crucial in our case.
For readers who are not familiar 
with this technique we give a discussion of this method in the 
appendix.
We thus present a 
fully quantum version of the Witten-Olive analysis.
The deformation of the supersymmetry algebra by
anomalies is such that the BPS bound remains saturated due to  
uniform shifts in energy and central charge.
In our algebraic approach, 
we initially formulate all results in terms of a total 
superfield, and do not decompose this superfield into a 
background and a quantum part. Only at the end do we need to 
use some properties of the kink background, namely the fact that 
the vacuum is annihilated by one of the supersymmetry 
charges (the time-independent charge).  

We would like to  summarize our results briefly.
It is  shown that the Noether current for ordinary 
(nonconformal) supersymmetry
\begin{equation}
j^{\mu}(x)
=-[\delslash\varphi+U(\varphi)]\gamma^{\mu}\psi(x)
\end{equation}
with  $U(\varphi)=g(\varphi^{2}-v^{2}_{0})$ 
contains in the present  formulation an apparent anomaly
\begin{equation}
\partial_{\mu}j^{\mu}(x)=\frac{\hbar g}{2\pi}\delslash\psi(x).
\end{equation}
(Only mass 
renormalization is necessary, $v_{0}^{2}=v^{2}+\delta v^{2}$,
and the mass counter term $\delta v^{2}$ is fixed by 
requiring that tadpoles vanish in the trivial 
vacuum~\cite{Rebhan:1997iv}.
For more general renormalization conditions, 
see~\cite{Rebhan:2002uk}).

As a result, the supercharge $Q^{\alpha}=\int dx j^{0,\alpha}$ 
is time dependent, though $\gamma_{\mu}j^{\mu}$ is shown to 
contain 
no anomalous component~\footnote{Divergences with respect to 
$\theta$ in superspace lead to anomalies in $x$-space which are 
of the form $\gamma_{\mu}j^{\mu}$ instead of 
$\partial_{\mu}j^{\mu}$. Similarly, the anomaly in the energy 
density is due to the trace anomaly, which itself is due to 
another $\theta$ divergence in superspace. From a superspace
point of view, $\theta$ divergences and $x$ divergences are 
equally fundamental.} but only an nonanomalous component 
\begin{equation}
(\gamma_{\mu}j^{\mu}(x))_{anomaly}=0.
\end{equation}
We thus define the conserved supercurrent 
\begin{equation}
\tilde{J}_{\mu}^{\alpha}(x)
=[j_{\mu}(x)
-\frac{\hbar g}{2\pi}\gamma_{\mu}\psi(x)]^{\alpha}; \ \ \ 
\partial_{\mu}\tilde{J}^{\mu}(x)=0
\end{equation}
and the associated time-independent supercharge
\begin{equation}
\tilde{Q}^{\alpha}=\int dx \tilde{J}^{0,\alpha}(x)
\end{equation}
which generates ordinary supersymmetry.
This conserved current contains the following anomalous
component
\begin{equation}
(\gamma_{\mu}\tilde{J}^{\mu}(x))_{anomaly}
=-\frac{\hbar g}{\pi}\psi(x).
\end{equation}
Because the model we consider breaks superconformal symmetry 
explicitly,
there is also a nonanomalous contribution to 
$\gamma_{\mu}\tilde{J}^{\mu}$ both at the tree and at the loop 
level. 

The supersymmetry algebra for the conserved charge $\tilde{Q}$
is then derived by using the BJL method. 
We rewrite Ward identities which are derived from
path integrals and which contain the covariant $T^{\star}$
time ordering, in terms of Ward identities at the operator 
level which contain the $T$ time ordering symbol. 
We obtain~\footnote{The symbols $(\gamma^{0})^{\alpha\beta}$ and
$(\gamma_{5})^{\alpha\beta}$ denote the matrices 
$(\gamma^{0})^{\alpha}_{\ \gamma}(C^{-1})^{\gamma\beta}$ and
$(\gamma_{5})^{\alpha}_{\ \gamma}(C^{-1})^{\gamma\beta}$ and 
are in our conventions equal to $-i$ and $i\tau_{3}$, 
respectively,
see Section 2. The indices $\alpha$ and $\beta$ are equal to 
$+$ or $-$, and for $\alpha=\beta=+$ one finds that the quantum 
anticommutator has the same form as the classical relation 
(1.1). }
\begin{eqnarray}
i\{\tilde{Q}^{\alpha},\tilde{Q}^{\beta}\}
&=&-2(\gamma^{\mu})^{\alpha\beta}\tilde{P}_{\mu}
-2\tilde{Z}(\gamma_{5})^{\alpha\beta}
\end{eqnarray}
where 
\begin{eqnarray}
&&\tilde{P}_{\mu}=\int dx \tilde{T}_{0\mu}(x), 
\ \ \ \tilde{H}=\tilde{P}_{0} ,\nonumber\\
&&\tilde{Z}=\int dx \tilde{\zeta}_{0}(x)
=-\int dx\tilde{\zeta}^{0}(x).
\end{eqnarray}

The BJL method, unlike the semi-classical Dirac bracket,
incorporates all the quantum effects, in particular 
superconformal anomalies.
The operators $\tilde{T}_{\nu}^{\ \mu}$ and $\tilde{\zeta}^{\mu}$
are conserved quantities
\begin{equation}
\partial_{\mu}\tilde{T}_{\nu}^{\ \mu}=0,\ \ \ \ \ 
\partial_{\mu}\tilde{\zeta}^{\mu}=0
\end{equation}
but contain superconformal anomalies
\begin{eqnarray}
\tilde{T}_{\mu}^{\ \mu}(x)&=&F(x)U(x)
-g\varphi\bar{\psi}\psi(x)+\frac{\hbar g}{2\pi}F(x)
\nonumber\\
&=& T_{\mu}^{\ \mu}(x)+\frac{\hbar g}{2\pi}F(x),
\nonumber\\
\gamma_{\mu}\tilde{\zeta}^{\mu}(x)\gamma_{5}
&=&\partial_{\mu}\varphi(x)U\gamma^{\mu}
+\frac{\hbar g}{2\pi}
\partial_{\mu}\varphi(x)\gamma^{\mu}
\nonumber\\
&=&(\gamma_{\mu}\zeta^{\mu}(x))\gamma_{5}
+\frac{\hbar g}{2\pi}
\partial_{\mu}\varphi(x)\gamma^{\mu}.
\end{eqnarray}
We derive these equations from the path integral
formulation.
In these equations, $T_{\mu}^{\ \mu}(x)$ and 
$(\gamma_{\mu}\zeta^{\mu})$ contain only the terms which 
explicitly break superconformal symmetry, as we shall show.
These arise from the superpotential, are ``soft'' (they have 
lower dimension because they are proportional to the 
dimensionful $g$),  and  there are no
anomalous contributions to these quantities.

The relations in (1.5) and (1.6) can be combined to give
a similar result as in (1.12)
\begin{equation} 
\gamma^{\mu}\tilde{J}_{\mu}(x)=\gamma^{\mu}j_{\mu}(x)
-\frac{\hbar g}{\pi}\psi(x).
\end{equation}
The other conserved operators themselves can also be decomposed 
in a similar way as in (1.6)
\begin{eqnarray}\label{1.13}
\tilde{T}_{\mu\nu}(x)&=&T_{\mu\nu}(x)+\eta_{\mu\nu}\frac{\hbar g}
{4\pi}F(x),\nonumber\\
\tilde{\zeta}_{\mu}(x)&=&\zeta_{\mu}(x)+\frac{\hbar g}
{2\pi}\epsilon_{\mu\nu}\partial^{\nu}\varphi(x).
\end{eqnarray}
The operators $T_{\mu\nu}(x)$ and $\zeta_{\mu}(x)$ are 
specified by 
\begin{equation}
T_{\mu}^{\ \mu}(x)_{anomaly}=0, \ \ \ \ \ 
\zeta^{\mu}(x)_{anomaly}=0
\end{equation}
corresponding to the absence of an anomaly in 
$\gamma_{\mu}j^{\mu}$, see (1.5),
 but although both $\tilde{\zeta}_{\mu}$ and $\zeta_{\mu}$ are
are conserved, $T_{\mu}^{\ \nu}$ is not conserved
\begin{equation}
\partial_{\nu}T_{\mu}^{\ \nu}(x)\neq 0,
\end{equation}
similar to the non-conservation of $j^{\mu}$.

In terms of $T_{\mu}^{\ \nu}$ and $\zeta^{\mu}$,
the supersymmetry algebra reads
\begin{eqnarray}
i\{\tilde{Q}^{\alpha},\tilde{Q}^{\beta}\}
&=&-2(\gamma^{\mu})^{\alpha\beta}P_{\mu}
+\int dx\frac{\hbar g}{2\pi}F(x)
(\gamma^{0})^{\alpha\beta}
\nonumber\\
&&-2Z(\gamma_{5})^{\alpha\beta}
+\int dx \frac{\hbar g}{\pi}\partial_{1}\varphi(x)(\gamma_{5})
^{\alpha\beta}
\end{eqnarray}
where we used $\epsilon^{01}=1$ and
\begin{eqnarray}
&&P_{\mu}=\int dx T_{0\mu}(x), 
\ \ \ H=P_{0} ,\nonumber\\
&&Z=\int dx \zeta_{0}(x)
=-\int dx\zeta^{0}(x).
\end{eqnarray}

We see that the supersymmetry algebra in terms of 
time-independent charges has the same form at the quantum level 
as at the classical level, see (1.10). This agrees 
with~ \cite{Shifman:1998zy}, whose analysis is based on this 
observation.
In (1.16) we have used charges $P$ and $Z$; $Z$ is free 
from anomalies and the anomaly of $\tilde{Z}$  explicitly 
appears on the right-hand side, but $P$ still contains 
a superconformal anomaly though $T_{\mu}^{\mu}$ is free 
from the trace anomaly. (In other words, $T_{00}$ and $T_{11}$
have equal anomalies, and the anomaly in $T_{00}$ doubles the 
contribution in~(\ref{1.13}) proportional to $F$, see Section 6.
Using (1.6) to write $\tilde{Q}$ in terms of $Q$, all the 
anomaly terms in (1.17) cancel separately if one splits off the
anomaly from $P_{\mu}$. In this way, also in terms of $Q$, 
$P_{\mu}$ and $Z$ the quantum anticommutator has the same form 
as classically. However, $Q$ and $P$ are time-dependent, so 
they are pysically less relevant.)       
These two alternative ways of writing the algebra give rise to
the same physical conclusion, namely, uniform shifts 
in energy and central charge in the vacuum of the time 
independent kink solution. Both maintain the BPS bound, since
they describe the same algebra on a different basis.

\section{The model and the superspace regulator }

We briefly summarize some of the features of the model which 
describes the supersymmetric model; the $N=(1,1)$  
supersymmetric 
Wess-Zumino model in $d=2$ Minkowski 
space. The 
model is defined in terms of the superfield
\begin{equation}
\phi(x,\theta)=\varphi(x)+\bar{\theta}\psi(x)
+\frac{1}{2}\bar{\theta}\theta F(x)
\end{equation}
where $\theta^{\alpha}$ is a Grassmann number, and 
$\theta^{\alpha}$ and $\psi^{\alpha}(x)$ are two-component
Majorana spinors; 
$\varphi(x)$ is a real 
scalar field, and $F(x)$ is a real auxiliary field. 
We define $\bar{\theta}=\theta^{T}C$ with $C$ the charge 
conjugation matrix, and the inner product for spinors is defined
 by 
\begin{equation}
\bar{\theta}\theta\equiv \theta^{T}C\theta = \theta^{\alpha}
C_{\alpha\beta}\theta^{\beta}\equiv
\bar{\theta}_{\beta}\theta^{\beta}
\end{equation}
with the Dirac  matrix convention
\begin{equation}
\gamma^{0}=-\gamma_{0}=-i\tau^{2}, \ \ \ \gamma^{1}=\gamma_{1}=
\tau^{3},\ \ \ 
C=\tau^{2}, \ \ \ \gamma_{5}=\gamma^{0}\gamma^{1}
\end{equation}
The choice $\gamma^{1}=\tau^{3}$ has certain advantages for the 
evaluation of the spectrum of the fermions; we shall not evaluate
this spectrum, but still use $\gamma^{1}=\tau^{3}$ in order to
agree with the literature.
We use the metric $\eta_{\mu\nu}=(-1,1)$ for $\mu=(0,1)$, hence
$(\gamma^{0})^{2}=-1$ but $\gamma_{5}^{2}=1$. 
Useful identities are $\epsilon^{\nu\mu}\gamma_{\mu}\gamma_{5}
=-\gamma^{\nu}$ and 
$\gamma^{\mu}\gamma_{5}=-\epsilon^{\mu\nu}\gamma_{\nu}$ with
$\epsilon^{01}=1$. Since in this representation the charge 
conjugation matrix equals $\tau^{2}$, the Majorana condition 
$\psi^{\dagger}\tau^{2}=\psi^{T}C$ reduces to the statement 
that all Majorana spinors are real. We frequently use the 
relations $\bar{\epsilon}\psi=\bar{\psi}\epsilon$, 
$\bar{\epsilon}\gamma^{\mu}\psi=-\bar{\psi}\gamma^{\mu}\epsilon$
and $\bar{\epsilon}\gamma_{5}\psi=-\bar{\psi}\gamma_{5}\epsilon$
(the sign in the last relation is opposite to the 4 dimensional 
case). 

The supersymmetry transformation is induced by its action 
on the coordinates in superspace, $\phi'(x',\theta')=\phi(x,\theta)$. One has 
\begin{eqnarray}
&&\theta'=\theta-\epsilon,\nonumber\\
&&x'^{\mu}= x^{\mu}-\bar{\theta}\gamma^{\mu}\epsilon 
\end{eqnarray}
leading in first order of $\epsilon$ to  
\begin{eqnarray}
&&\phi'(x,\theta)
=\phi(x^{\mu}+\bar{\theta}\gamma^{\mu}\epsilon,\theta+\epsilon)
=\nonumber\\
&&\ \ \ \ \ \ \ \ \ \ \ \ \ \ \  \phi(x,\theta)
+ \bar{\theta}\gamma^{\mu}\epsilon
\partial_{\mu}\varphi(x)+\bar{\epsilon}\psi(x)
+\bar{\theta}\epsilon F(x)+\frac{1}{2}(\bar{\theta}\theta)
\bar{\epsilon}\gamma^{\mu}\partial_{\mu}\psi
\end{eqnarray}
where we used a Fierz rearrangement
\begin{equation}
(\bar{\theta}\partial_{\mu}\psi)
(\bar{\theta}\gamma^{\mu}\epsilon)
=\frac{1}{2}(\bar{\theta}\theta)
\bar{\epsilon}\gamma^{\mu}\partial_{\mu}\psi.
\end{equation}
In terms of components one obtains from $\delta\phi=\phi'(x,\theta)-\phi(x,\theta)$
\begin{eqnarray}
&&\delta\varphi=\bar{\epsilon}\psi(x),\nonumber\\
&&\delta\psi=\partial_{\mu}\varphi(x)\gamma^{\mu}\epsilon
+F(x)\epsilon=\delslash\varphi(x)\epsilon + F(x)\epsilon,
\nonumber\\
&&\delta F=\bar{\epsilon}\gamma^{\mu}\partial_{\mu}\psi
=\bar{\epsilon}\delslash\psi(x).
\end{eqnarray}
The supercharge which generates (2.4)
\begin{eqnarray}
&&\bar{\epsilon}Q\equiv \bar{\epsilon}_{\alpha}
\frac{\partial}{\partial\bar{\theta}_{\alpha}}
-\bar{\epsilon}\gamma^{\mu}\theta\partial_{\mu}
\end{eqnarray}
and the covariant derivative 
\begin{eqnarray}
&&\bar{\eta}D\equiv \bar{\eta}_{\alpha}
\frac{\partial}{\partial\bar{\theta}_{\alpha}}
+\bar{\eta}\gamma^{\mu}\theta\partial_{\mu}\nonumber  
\end{eqnarray}
anti-commute with each other.
We have
\begin{eqnarray}
D^{\alpha}\phi(x,\theta)
&=&\psi^{\alpha}+\theta^{\alpha}F
+(\gamma^{\mu}\theta)^{\alpha}\partial_{\mu}\varphi
+(\gamma^{\mu}\theta)^{\alpha}\bar{\theta}\partial_{\mu}\psi
\end{eqnarray}
and 
\begin{eqnarray}
\bar{D}\phi(x,\theta)
D\phi(x,\theta)
&=&\bar{\psi}\psi+2\bar{\psi}\theta F
+2(\bar{\psi}\gamma^{\mu}\theta)\partial_{\mu}\varphi
\nonumber\\
&&+\bar{\theta}\theta[FF
-\partial^{\mu}\varphi\partial_{\mu}\varphi
-\bar{\psi}\gamma^{\mu}\partial_{\mu}\psi]
\end{eqnarray}
where we used $\bar{\theta}\gamma^{\mu}\theta=0$.

We next note that 
\begin{eqnarray}
&&\phi^{3}(x,\theta)=\varphi^{3}+3(\bar{\theta}\psi)\varphi^{2}
+\frac{1}{2}(\bar{\theta}\theta)[3F\varphi^{2}
-3(\bar{\psi}\psi)\varphi]
\end{eqnarray}
by using $(\bar{\theta}\psi)(\bar{\theta}\psi)=
-\frac{1}{2}(\bar{\theta}\theta)(\bar{\psi}\psi)$.
We thus choose the action
\begin{eqnarray}
&&\int dx d^{2}\theta {\cal L}(x,\theta)=\int dxd^{2}\theta 
[\frac{1}{4}\bar{D}\phi(x,\theta)D\phi(x,\theta)
+\frac{1}{3}g\phi^{3}(x,\theta)
-gv^{2}_{0}\phi(x,\theta)]\nonumber\\
&&=\int dx\{\frac{1}{2}[FF
-\partial^{\mu}\varphi\partial_{\mu}\varphi
-\bar{\psi}\gamma^{\mu}\partial_{\mu}\psi]
+gF\varphi^{2}-g(\bar{\psi}\psi)\varphi
-gv^{2}_{0}F\}
\end{eqnarray}
where we used the convention
\begin{equation}
\int d^{2}\theta \frac{1}{2}(\bar{\theta}\theta)=1
\end{equation}
The delta function is defined by 
$\int d^{2}\theta_{1}\delta(\theta_{1}-\theta_{2})=1$
and  given by
\begin{equation}
\delta(\theta_{1}-\theta_{2})
=\frac{1}{2}\overline{(\theta_{1}-\theta_{2})}
(\theta_{1}-\theta_{2}).
\end{equation}
The potential $V$ in ${\cal L}=T-V$ is given by 
$V=-FU+g\bar{\psi}\psi\varphi$ where  
\begin{equation}
U(\varphi)\equiv g(\varphi^{2}-v^{2}_{0}).
\end{equation}
We use a coupling constant $g$ which is related to the coupling
constant $\lambda$ used in other articles by 
\begin{equation}
g=\sqrt{\frac{\lambda}{2}}, \ \ v_{0}
=\frac{\mu_{0}}{\sqrt{\lambda}}=\frac{m_{0}}{2g}.
\end{equation}
This coupling constant $g$ has the dimension of a mass, and 
$\mu_{0}^{2}=\mu^{2}+\Delta\mu^{2}$ where $m=\sqrt{2}\mu$ is 
the renormalized meson mass.

To apply the background field method, we decompose the field
variable as follows
\begin{equation}
\phi(x,\theta)=\Phi(x,\theta)+\eta(x,\theta)
\end{equation}
where $\Phi(x,\theta)$ is the background field and 
$\eta(x,\theta)$ is the quantum fluctuation. We then consider 
the parts of the (superfield) Lagrangian which are quadratic in 
$\eta$
\begin{equation}
{\cal L}_{2}(x,\theta)=\frac{1}{4}\bar{D}\eta(x,\theta)
D\eta(x,\theta)+g\Phi(x,\theta)\eta^{2}(x,\theta)
\end{equation}
or equivalently
\begin{eqnarray}
{\cal L}_{2}(x,\theta)&=&\eta(x,\theta)\Gamma(x,\theta)
\eta(x,\theta),\nonumber\\
\Gamma(x,\theta)&=&-\frac{1}{4}\bar{D}D
+g\Phi(x,\theta).
\end{eqnarray}

The regulator will be quadratic in $\Gamma$. To construct it,
we first derive some basic properties of the operator 
$\bar{D}D$. From the definition~\footnote{The symbol $T^{\star}$
denotes (covariant) time ordering in the path integral approach.
It has the property that it commuters with ordinary derivatives 
$\frac{\partial}{\partial x^{\mu}}$.}
\begin{equation}
\langle 0|T^{\star} \eta(x^{\mu},\theta_{1})
\eta(y^{\mu},\theta_{2})
|0\rangle
=-iG(x-y; \theta_{1},\theta_{2})
\end{equation}
and for a supersymmetric vacuum, we have
\begin{equation}
\langle0| T^{\star} \eta(x^{\mu}-\bar{\theta}_{1}\gamma^{\mu}\epsilon,
\theta_{1}-\epsilon)
\eta(y^{\mu}-\bar{\theta}_{2}\gamma^{\mu}\epsilon,
\theta_{2}-\epsilon)|0\rangle
=-iG(x-y; \theta_{1},\theta_{2}).
\end{equation}
In particular for $\epsilon=\theta_{2}$, by using 
$\bar{\theta}_{2}\gamma^{\mu}\theta_{2}=0$ and 
$\bar{\epsilon}\gamma^{\mu}\epsilon=0$, we obtain
\begin{equation}
\langle0| T^{\star} \eta(x^{\mu}-\bar{\theta}_{1}\gamma^{\mu}
\theta_{2}, \theta_{1}-\theta_{2})
\eta(y^{\mu},0)|0\rangle
=-iG(x-y; \theta_{1},\theta_{2}).
\end{equation}
For the massless theory there are no off-diagonal terms in the 
 propagator, hence the {\em free} field propagator in 
superspace can be written in terms of   
 the propagator of the scalar field $\varphi$
\begin{equation}
G(x-y; \theta_{1},\theta_{2})=\int\frac{d^{2}k}{(2\pi)^{2}}
e^{ik(x-y)}\frac{\exp[-i\bar{\theta}_{1}\kslash\theta_{2}]}
{k^{2}-i\epsilon}
\end{equation}
with $k^{2}=k^{2}_{1}-k^{2}_{0}$. This shows 
that~\footnote{ Completing squares in 
$
Z=\int d\mu  \exp\frac{i}{\hbar}[-\frac{1}{4}\eta\bar{D}D\eta
+J\eta]$
by setting $\eta=(\frac{1}{2}\bar{D}D)^{-1}J$ yields 
$Z=\int d\mu  \exp\frac{i}{\hbar}J(\bar{D}D)^{-1}J]$. 
Differentiation with respect to $J$ yields 
$\langle T^{\star}\eta(x,\theta_{1})\eta(y,\theta_{2})\rangle
=-i\hbar 
(\frac{1}{2}\bar{D}D)^{-1}
\delta(x-y)\delta(\theta_{1}-\theta_{2})$. We set $\hbar=1$
in most places.}
\begin{eqnarray}
&&G(x-y; \theta_{1},\theta_{2})=
\frac{\exp[-i\bar{\theta}_{1}\pslash\theta_{2}]}
{p^{2}-i\epsilon}\delta(x-y),\nonumber\\
&&\frac{1}{2}\bar{D}DG(x-y; \theta_{1},\theta_{2})=
\delta(x-y)\delta(\theta_{1}-\theta_{2})
\end{eqnarray}
where
\begin{equation}
p_{\mu}=\frac{1}{i}\partial_{\mu}.
\end{equation}
This equation yields the following important relations
\begin{eqnarray}
&&\frac{1}{2}\bar{D}D
\frac{\exp[-i\bar{\theta_{1}}\pslash\theta_{2}]}{p^{2}-i\epsilon}
=\delta(\theta_{1}-\theta_{2}),\nonumber\\
&&-\frac{1}{2}\bar{D}D\delta(\theta_{1}-\theta_{2})
=\exp[-i\bar{\theta_{1}}\pslash\theta_{2}],\nonumber\\
&&\exp[-i\bar{\theta_{1}}\pslash\theta_{2}]=1
-i\bar{\theta_{1}}\pslash\theta_{2}-p^{2}\delta(\theta_{1})
\delta(\theta_{2})
\end{eqnarray}
where the second relation is derived by performing the integral 
$\int d^{2}\theta_{2}$ in
\begin{equation}
\frac{1}{2}\bar{D}D
\frac{\exp[-i\overline{(\theta_{1}-\theta_{3})}
\pslash\theta_{2}]}
{p^{2}-i\epsilon}
=\delta(\theta_{1}-\theta_{2})\exp[i\bar{\theta}_{3}
\pslash\theta_{2}].
\end{equation}
We thus have
\begin{eqnarray}\label{2.28}
&&{\cal D}\equiv \frac{1}{2}\bar{D}D,\nonumber\\
&&{\cal D}(\theta_{1})\delta(\theta_{1}-\theta_{2})
=-\exp[-i\bar{\theta_{1}}\pslash\theta_{2}],\nonumber\\
&&{\cal D}^{2}(\theta_{1})\delta(\theta_{1}-\theta_{2})
=-p^{2}\delta(\theta_{1}-\theta_{2}).
\end{eqnarray}
Using these results, we obtain the following equations 
\begin{eqnarray}
&&\Gamma(x, \theta)=-\frac{1}{2}{\cal D}+g\Phi(x,\theta),
\nonumber\\
&&\Gamma^{2}(x, \theta)=\frac{1}{4}{\cal D}^{2}-\frac{1}{2}g{\cal D}\Phi
-\frac{1}{2}g\Phi{\cal D}+(g\Phi)^{2},\nonumber\\
&&\Gamma^{2}(x, \theta)\delta(\theta-\theta_{1})
=[-\frac{1}{4}p^{2}-\frac{1}{2}g{\cal D}\Phi
-\frac{1}{2}g\Phi{\cal D}+(g\Phi)^{2}]\delta(\theta-\theta_{1})
,\nonumber\\
&&[\Gamma^{2}(x, \theta)]^{2}\delta(\theta-\theta_{1})
=\int d^{2}\theta_{2}\Gamma^{2}(x, \theta)\delta(\theta-\theta_{2})
\Gamma^{2}(x, \theta_2)\delta(\theta_{2}-\theta_{1})
\nonumber\\
&&=[-\frac{1}{4}p^{2}-\frac{1}{2}g{\cal D}\Phi
-\frac{1}{2}g\Phi{\cal D}+(g\Phi)^{2}]^{2}
\delta(\theta-\theta_{1}).
\end{eqnarray}

We use the heat kernel $\exp[(\Gamma/M)^{2}]$ as regulator in superspace.
Using (2.29) we find 
\begin{eqnarray}
&&(\exp[\frac{1}{M^{2}}\Gamma(x,\theta)^{2}])
\delta(\theta-\theta_{1})\nonumber\\
&&=(\exp\{\frac{1}{M^{2}}[-\frac{1}{4}p^{2}
-\frac{1}{2}g{\cal D}\Phi
-\frac{1}{2}g\Phi{\cal D}+(g\Phi)^{2}]\})
\delta(\theta-\theta_{1}).
\end{eqnarray}
This last equation will be used to evaluate the Jacobian.

\section{Evaluation of the Jacobian}
The regularized Jacobian $J$ in the path integral
\begin{equation}
\int{\cal D}\phi\exp[i\int d^{2}x d^{2}\theta{\cal L}(x,\theta)]
\end{equation}
for the infinitesimal transformation
\begin{equation}
\phi^{\prime}(x,\theta)=\phi(x,\theta)
+\omega(x,\theta)\phi(x,\theta)
\end{equation}
is given at the one-loop level by
\begin{eqnarray}
&&\ln J
=\int d^{2}x d^{2}\theta 
\langle x,\theta|\omega(x,\theta)
\exp\{\frac{1}{M^{2}}\Gamma(x,\theta)^{2}
\}|\theta,x\rangle\nonumber\\
&=&\int d^{2}x d^{2}\theta\omega^{\prime}(x,\theta)
\langle x,\theta|
\exp\{\frac{1}{M^{2}}[-\frac{1}{2}{\cal D}+g\Phi(x,\theta)]^{2}
\}|\theta,x\rangle\nonumber\\
&=&\int d^{2}x d^{2}\theta\omega^{\prime}(x,\theta)
\nonumber\\
&&\times
\lim_{y\rightarrow x,\theta_{1}\rightarrow\theta}
\exp\{\frac{1}{M^{2}}[-\frac{1}{2}{\cal D}+g\Phi(x,\theta)]^{2}
\}
\delta(\theta-\theta_{1})
\delta(x-y)\\
&=&
\int d^{2}x d^{2}\theta\frac{d^{2}k}{(2\pi)^{2}} 
\omega^{\prime}(x,\theta)e^{-ikx}
\nonumber\\
&&\times\lim_{\theta_{1}\rightarrow\theta} 
\exp\{\frac{1}{M^{2}}[\frac{1}{4}{\cal D}^{2}
-\frac{1}{2}{\cal D}g\Phi(x,\theta)-
\frac{1}{2}g\Phi(x,\theta){\cal D}+g^{2}\Phi^{2}]\}e^{ikx}
\delta(\theta-\theta_{1})
\nonumber\\
&=&
\int d^{2}x d^{2}\theta\frac{d^{2}k}{(2\pi)^{2}} 
\omega^{\prime}(x,\theta)e^{-ikx}
\nonumber\\
&&\times\lim_{\theta_{1}\rightarrow\theta} 
\exp\{\frac{1}{M^{2}}[-\frac{1}{4}p^{2}-\frac{1}{2}
{\cal D}g\Phi(x,\theta)-
\frac{1}{2}g\Phi(x,\theta){\cal D}+g^{2}\Phi^{2}]\}e^{ikx}
\delta(\theta-\theta_{1}).
\nonumber
\end{eqnarray}
We recall
\begin{eqnarray}
&&{\cal D}=\frac{1}{2}\bar{D}D,\nonumber\\
&&p_{\mu}=(1/i)\partial_{\mu}
\end{eqnarray}
and the parameter $M$ is sent to $\infty$ later. 

When one 
considers a transformation of the form
\begin{eqnarray}
\phi^{\prime}(x,\theta)&=&\phi(x,\omega)
+[\bar{\Omega}(x,\theta)D\phi(x,\theta)
+\bar{\epsilon}(x,\theta)Q\phi(x,\theta)
+O(x,\theta)]\phi(x,\theta)\nonumber\\
&\equiv&\phi(x,\theta)+\omega(x,\theta)\phi
\end{eqnarray}
the parameter $\omega(x,\theta)$ in the first line in (3.3)
must be replaced by $\omega^{\prime}(x,\theta)$ in the second 
line where
\begin{equation}
\omega^{\prime}(x,\theta)=
\frac{1}{2}\bar{\Omega}(x,\theta)D
+\frac{1}{2}\bar{\epsilon}(x,\theta)Q
+O(x,\theta).
\end{equation}
The reason for the factor $1/2$ is that when one extracts the 
factor $\omega(x,\theta)$ outside the 
bracket symbol, the derivative operators 
$D^{\alpha}$  and $Q^{\alpha}$ ( and also the spatial 
derivative $\partial_{\mu}$) act on both the 
bra- and ket- vectors.
\\

We start with the evaluation of
\begin{equation}
\int\frac{d^{2}k}{(2\pi)^{2}} 
e^{-ikx} 
\exp\{\frac{1}{M^{2}}[-\frac{1}{4}p^{2}-\frac{1}{2}
{\cal D}g\Phi(x,\theta)-
\frac{1}{2}g\Phi(x,\theta){\cal D}+g^{2}\Phi^{2}]\}e^{ikx}
\delta(\theta-\theta_{1}).
\end{equation}
Passing the factor 
$e^{ikx}$ through the integrand replaces $p\rightarrow
p+k$, and the operator ${\cal D}$ is modified as follows
\begin{eqnarray}
e^{-ikx}{\cal D}e^{ikx}&=&
\frac{1}{2}[\bar{D}+i\overline{(\gamma^{\nu}\theta)}k_{\nu}]
[D+i(\gamma^{\mu}\theta)k_{\mu}]
\nonumber\\
&=&\frac{1}{2}\bar{D}D+i\overline{(\kslash\theta)}D
-\frac{1}{2}\overline{(\kslash\theta)}(\kslash\theta)
\nonumber\\
&=&{\cal D}
-i\bar{\theta}\kslash D+\frac{1}{2}k^{2}(\bar{\theta}\theta).
\end{eqnarray}
Replacing
\begin{equation}
k_{\mu}\rightarrow Mk_{\mu}
\end{equation}
we obtain the integral
\begin{eqnarray}
&&M^{2}\int\frac{d^{2}k}{(2\pi)^{2}}  
(\exp\{-\frac{1}{4}(k^{2}+2\frac{kp}{M}+\frac{p^{2}}{M^{2}})
-\frac{1}{2}[\frac{{\cal D}}{M^{2}}
-i\frac{\bar{\theta}\kslash D}{M}+\frac{1}{2}k^{2}
(\bar{\theta}\theta)]
g\Phi(x,\theta)\nonumber\\
&&\ \ \ \ -\frac{1}{2}g\Phi(x,\theta)[\frac{{\cal D}}{M^{2}}
-i\frac{\bar{\theta}\kslash D}{M}+\frac{1}{2}k^{2}
(\bar{\theta}\theta)]
+\frac{g^{2}}{M^{2}}\Phi^{2}\})
\delta(\theta-\theta_{1}).
\end{eqnarray}
We now expand the exponent into a power series
except for the factor $\exp[-\frac{1}{4}k^{2}]$. By using that according to~(\ref{2.28})
\begin{equation}
\lim_{\theta_{1}\rightarrow\theta}{\cal D}(\theta)
\delta(\theta-\theta_{1})=-1
\end{equation}
while  terms without ${\cal D}$ acting on 
$\delta(\theta-\theta_{1})$ vanish for 
$\theta_{1}\rightarrow\theta$, 
and noting that only 
the terms in the integral of order $1/M^{2}$ 
or larger survive when $M$ tends to infinity, one can confirm 
that only the terms to second order in the expansion survive. 
In fact, the second order terms completely cancel because the 
term
\begin{eqnarray}
k^{2}(\bar{\theta}\theta){\cal D}/M^{2}
\end{eqnarray}
from the cross terms of ${\cal D}/M^{2}$ and $\frac{1}{2}k^{2}
(\bar{\theta}\theta)$ cancels the square
\begin{equation}
[-i\bar{\theta}\kslash D]^{2}/M^{2}=
-\frac{1}{2}(\bar{\theta}\theta)
k^{2}\bar{D}D/M^{2}=-k^{2}(\bar{\theta}\theta){\cal D}/M^{2}.
\end{equation}

We thus need to evaluate only the first order terms
\begin{eqnarray}
\int\frac{d^{2}k}{(2\pi)^{2}}  
\exp\{-\frac{1}{4}k^{2}\}[-g\Phi(x,\theta){\cal D}]
\delta(\theta-\theta_{1})=i\frac{g}{\pi}\Phi(x,\theta).
\end{eqnarray}
We used the  formula
\begin{eqnarray}
&&\int\frac{d^{2}k}{(2\pi)^{2}} \exp[-k^{2}/4]=\frac{i}{\pi}
\end{eqnarray}
where in the last integral we Wick-rotated to Euclidean space 
by $d^{2}k\rightarrow id^{2}k$.
We thus obtain the following result for the Jacobian in 
Minkowski space
\begin{equation}
\ln J=i\int d^{2}xd^{2}\theta\omega^{\prime}(x,\theta)
[\frac{g}{\pi}\Phi(x,\theta)].
\end{equation}
Note that this calculation remains valid for general 
(non-derivative) interactions depending on $\Phi$; if one has 
a potential $V(\Phi)$ instead of $g\Phi$ in (2.19), one 
makes the same replacement in (3.16).

In the spirit of the background field method, one may replace 
the 
variable $\Phi(x,\theta)$ by the full variable $\phi(x,\theta)$
to the accuracy of the one-loop approximation.
The final  result for the result of the Jacobian of the path 
integral in Minkowski space is thus given by
\begin{equation}
\ln J=i\int d^{2}xd^{2}\theta\omega^{\prime}(x,\theta)
[\frac{ g}{\pi}\phi(x,\theta)].
\end{equation}
For example, for the class of transformations
\begin{equation}
\delta\phi(x,\theta)=\bar{\Omega}D\phi+c(\bar{D}\Omega)\phi
=\bar{\Omega}D\phi+\frac{1}{2}(\bar{D}\Omega)\phi
+(c-\frac{1}{2})(\bar{D}\Omega)\phi
\end{equation}
with a constant $c$, one obtains for the Jacobian
\begin{eqnarray}
&&i\frac{g}{\pi}\int d^{2}xd^{2}\theta
\left[ \frac{1}{2}\bar{D}(\Omega
\Phi(x,\theta))+ (c-\frac{1}{2})\left(\bar{D}\Omega(x,\theta)\right)\right]
\phi(x,\theta)\nonumber\\
&&=-i(c-\frac{1}{2})\frac{g}{\pi}\int d^{2}xd^{2}\theta
\bar{\Omega}(x,\theta)D\phi(x,\theta).
\end{eqnarray}
We emphasize that this is a one-loop result.

\section{Superconformal anomalies from the measure}

Our analysis of superconformal anomalies starts with
the generalized supersymmetry transformation
\begin{equation}\label{4.1}
\delta\phi(x,\theta)=\bar{\Omega}(x,\theta)Q\phi(x,\theta)
\end{equation}
which gives rise to a change of the  action
\begin{equation}
\delta S=\int d^{2}xd^{2}\theta [\frac{1}{2}(D^{\alpha}
\bar{\Omega}_{\beta})(\bar{D}_{\alpha}\phi) Q^{\beta}\phi
+\bar{\Omega}_{\alpha}Q^{\alpha}{\cal L}]
\end{equation}
where
\begin{equation}
{\cal L}=\frac{1}{4}\bar{D}_{\alpha}\phi D^{\alpha}
\phi+\frac{1}{3}g\phi^{3}-gv^{2}_{0}\phi.
\end{equation}
Any transformation of $\phi$, whether it is a symmetry of the 
action or not, leads to a corresponding Ward identity, but 
using a local supersymmetry transformation has the advantage that
one obtains a hierarchy of Ward identitites in $x$-space which 
contain the Ward identities for ordinary and conformal 
supersymmetry. These are, of course, the Ward identities we are 
interested in, and we expect in this multiplet of Ward 
identities also to find a Ward identity for the central charge
current.   

For constant superfields $\Omega^{\alpha}$, the action is 
invariant, but
for local $\Omega^{\alpha}$, the variation of $S$ is 
porportional to the Noether current.
One thus obtains the following Ward identity for correlation
functions
\begin{eqnarray}
&&\langle\frac{i}{\hbar}\int d^{2}xd^{2}\theta 
[\frac{1}{2}(D^{\alpha}
\bar{\Omega}_{\beta})(\bar{D}_{\alpha}\phi) Q^{\beta}\phi
+\bar{\Omega}_{\alpha}Q^{\alpha}{\cal L}]\phi(x_{1},\theta_{1})
...\phi(x_{n},\theta_{n})\rangle\nonumber\\
&&=\langle-i\frac{g}{2\pi}\int d^{2}xd^{2}\theta 
\bar{\Omega}_{\beta}Q^{\beta}\phi(x,\theta)\phi(x_{1},\theta_{1})
...\phi(x_{n},\theta_{n})\rangle\nonumber\\ 
&&-\langle\delta\phi(x_{1},\theta_{1})
...\phi(x_{n},\theta_{n})\rangle  ...
-\langle\phi(x_{1},\theta_{1})
...\delta\phi(x_{n},\theta_{n})\rangle. 
\end{eqnarray}
The first term on the right-hand side is the Jacobian; it 
contains all anomalies, and the  extra factor of $1/2$ arises 
from the rule (3.6). We shall from now on replace Ward 
identities such as (4.4) by the simplified operator 
expression 
\begin{equation}
\frac{1}{2}D^{\alpha}[(\bar{D}_{\alpha}\phi) Q^{\beta}\phi]
+Q^{\beta}{\cal L}= -\frac{\hbar g}{2\pi}Q^{\beta}\phi(x,\theta).
\end{equation}
Here we write $\hbar$ explicitly to indicate
that we are working at the one-loop level. 

Using $\frac{1}{2}D^{\alpha}[(\bar{D}_{\alpha}\phi)
Q^{\beta}\phi] + Q^{\beta}{\cal L}=
\frac{1}{2}(D^{\alpha}\bar{D}_{\alpha}\phi) Q^{\beta}\phi
- Q^{\beta}V$, and $\frac{1}{2}D^{\alpha}\bar{D}_{\alpha}\phi=F
+\bar{\theta}~\!\!\!\!\not~\!\!\!\!\!\!~\partial\psi
+\frac{1}{2}\bar{\theta}\theta\partial_{\mu}\partial^{\mu}
\varphi$
and $Q^{\beta}\phi=\psi^{\beta}+F\theta^{\beta}-\delslash\varphi
\theta^{\beta}
+\frac{1}{2}\bar{\theta}\theta(\delslash\psi)^{\beta}$, 
the Ward identity in superspace can be expanded 
as follows
\begin{eqnarray}\label{4.6}
&&\frac{1}{2}(\gamma_{\mu}\tilde{J}^{\mu})(x)
-\tilde{T}_{\mu}^{\ \mu}(x)\theta 
+\frac{1}{2}(\bar{\psi}\gamma_{5}\delslash\psi)\gamma_{5}\theta
+\gamma_{\mu}\tilde{\zeta}^{\mu}(x)\gamma_{5}\theta
+\frac{1}{2}(\bar{\psi}\gamma^{\nu}\delslash\psi)\gamma_{\nu}
\theta\nonumber\\
&&
-\delta(\theta)\partial_{\mu}j^{\mu}(x)\nonumber\\
&&=\frac{1}{2}(\gamma_{\mu}j^{\mu})(x)
-(T_{\mu}^{\ \mu})(x)\theta
+(\gamma_{\mu}\zeta^{\mu})(x)\gamma_{5}\theta \nonumber\\
&&-\frac{\hbar g}{2\pi}[\psi(x)+F(x)\theta-(\gamma^{\mu}\theta)
\partial_{\mu}\varphi(x)+\delta(\theta)\delslash\psi(x)]
\end{eqnarray}
or in component notation
\begin{eqnarray}\label{4.7}
&&(\gamma_{\mu}\tilde{J}^{\mu})(x)=(\gamma_{\mu}j^{\mu})(x)
-\frac{\hbar g}{\pi}\psi(x),\nonumber\\
&&\tilde{T}_{\mu}^{\ \mu}(x)=(T_{\mu}^{\ \mu})(x)
+\frac{\hbar g}{2\pi}F(x),\nonumber\\
&&\tilde{\zeta}^{\mu}(x) 
-\frac{1}{2}(\bar{\psi}\gamma^{\mu}\gamma_{5}\delslash\psi)(x)
=(\zeta^{\mu})(x)
+\frac{\hbar g}{2\pi}
\epsilon^{\mu\sigma}\partial_{\sigma}\varphi(x),\nonumber\\
&&(\bar{\psi}\gamma_{5}\delslash\psi)(x)=0,\nonumber\\
&&\partial_{\mu}j^{\mu}(x)=\frac{\hbar g}{2\pi}\delslash\psi(x).
\end{eqnarray}
All superconformal anomalies are 
included. The operators $\gamma_{\mu}\tilde{J}^{\mu}$,
$\tilde{T}_{\mu}^{\ \mu}$ and $\tilde{\zeta}^{\mu}$ 
are constructed only from the part $\bar{D}\phi D\phi$ of the 
Lagrangian. This is the massless part which is superconformally 
invariant, and these terms appear on the left-hand side of~(\ref{4.6}) 
and~(\ref{4.7}). The terms coming from the superpotential depend on 
the dimensionful coupling constant $g$ and thus break 
superconformal invariance. These terms appear on the right-hand
side of~(\ref{4.6}) and~(\ref{4.7}), and they are given by 
\begin{eqnarray}
&&\frac{1}{2}(\gamma_{\mu}j^{\mu})(x)
-(T_{\mu}^{\ \mu})(x)\theta
+(\gamma_{\mu}\zeta^{\mu})(x)\gamma_{5}\theta
-\delta(\theta)\partial_{\mu}(U\gamma^{\mu}\psi)
\nonumber\\
&&\equiv QV(\phi(x,\theta))\nonumber\\
&&=-U(\varphi)\psi-[FU(\varphi)-g\varphi\bar{\psi}\psi]\theta
+\partial_{\mu}\varphi U(\varphi)\gamma^{\mu}\theta
-\delta(\theta)\partial_{\mu}(U\gamma^{\mu}\psi)
\end{eqnarray}
where 
$V=-[\frac{1}{3}g\phi^{3}(x,\theta)-gv^{2}_{0}\phi(x,\theta)]$ 
is the superpotential.
The term proportional to $\delta(\theta)$ in (4.8) has been
moved to
 the left-hand side of (4.6) where it forms  part of 
$-\delta(\theta)\partial_{\mu}j^{\mu}(x)$.

The Ward identity contains contractions of currents and does 
not specify the full expressions of 
the various operators involved.
The full expressions of the supercurrent $j^{\mu}$ 
(and $\tilde{J}^{\mu}$), energy-momentum
tensor $T_{\mu\nu}$ (and $\tilde{T}_{\mu\nu}$) and central charge
current $\zeta_{\mu}$ (and $\tilde{\zeta}_{\mu}$) are, respectively, given by
\begin{eqnarray}\label{4.9}
j^{\mu}(x)&=&-[\delslash\varphi(x)+U(\varphi(x))]\gamma^{\mu}
\psi(x),\nonumber\\
&&\nonumber\\
\tilde{J}^{\mu}(x)&=&-[\delslash\varphi(x)-F(x)]\gamma^{\mu}
\psi(x)\nonumber\\
&=&j^{\mu}-\frac{\hbar g}{2\pi}\gamma^{\mu}\psi,\nonumber\\
T_{\mu\nu}(x)&=&
\partial_{\mu}\varphi\partial_{\nu}\varphi
-\frac{1}{2}\eta_{\mu\nu}[(\partial^{\rho}\varphi)
(\partial_{\rho}\varphi)-FU]\nonumber\\
&+&\frac{1}{4}\bar{\psi}[\gamma_{\mu}\partial_{\nu}+\gamma_{\nu}
\partial_{\mu}]\psi
-\frac{1}{4}\eta_{\mu\nu}
[\bar{\psi}\gamma^{\rho}\partial_{\rho}\psi
+2g\varphi \bar{\psi}\psi],\nonumber\\
\tilde{T}_{\mu\nu}(x)&=&
\partial_{\mu}\varphi\partial_{\nu}\varphi
-\frac{1}{2}\eta_{\mu\nu}[(\partial^{\rho}\varphi)
(\partial_{\rho}\varphi)+F^{2}]
+\frac{1}{4}\bar{\psi}[\gamma_{\mu}\partial_{\nu}+\gamma_{\nu}
\partial_{\mu}]\psi\nonumber\\
&=&T_{\mu\nu}(x)
+\eta_{\mu\nu}\frac{\hbar g}{4\pi}F(x),
\nonumber\\
\zeta_{\mu}(x)&=&\epsilon_{\mu\nu}\partial^{\nu}\varphi(x)
U(\varphi), \nonumber\\
&&\nonumber\\
\tilde{\zeta}_{\mu}(x)&=&-\epsilon_{\mu\nu}
\partial^{\nu}\varphi(x)F(x)
\nonumber\\
&=&\zeta_{\mu}(x)+\frac12{\bar \psi}\gamma_{\mu}\gamma_5\delslash\psi(x)+\frac{\hbar g}{2\pi}\epsilon_{\mu\nu}
\partial^{\nu}\varphi(x)
\end{eqnarray}
with $U(\varphi)=g(\varphi^{2}(x)-v^{2}_{0})$ and 
$\epsilon_{01}=-1$. These expressions for the various operators 
are derived later. At this point we only know that they satisfy
(4.7). One may already note that the currents $j^{\mu}$ and 
$\tilde{J}^{\mu}$, $T_{\mu\nu}$ and
${\tilde T}_{\mu\nu}$, and $\zeta_{\mu}$ and ${\tilde \zeta}_{\mu}$ only differ by 
the $F$ and $\psi$ field equations. We repeat that these 
relations between $\tilde{J}_{\mu}$ and $j_{\mu}$,  
${\tilde T}_{\mu\nu}$ and $ T_{\mu\nu}$, and ${\tilde \zeta}_{\mu}$ and $\zeta_{\mu}$, are only 
identities when used in the Ward identity (4.4).

We now give some discussion of (4.7) and (4.9).
One may expand the spinor parameter $\Omega^{\alpha}$ in (4.1)
as follows
\begin{equation}\label{4.10}
\Omega^{\alpha}(x,\theta)=s^{\alpha}(x)+w(x)\theta^{\alpha} +
l(x)(\gamma_{5}\theta)^{\alpha} + 
c_{\mu}(x)(\gamma^{\mu}\gamma_{5}\theta)^{\alpha}
+t^{\alpha}(x)\delta(\theta).\nonumber
\end{equation}
The parameter $s^{\alpha}(x)$ generates local ordinary
supersymmetry
and thus yields a Ward identity for the divergence of the supercurrent, 
see the terms with $\delta(\theta)$ in (4.6).
The parameter $w(x)$ 
generates Weyl transformations in~(\ref{4.1}), namely 
$\delta\phi=w(x)\theta^{\alpha}
\frac{\partial}{\partial{\theta^{\alpha}}}\phi$, and thus 
yields the 
trace anomaly. The 
parameter $l(x)$ generates local Lorentz transformations (or
chiral transformations since $\gamma_{5}$ is the Lorentz 
generator), but the Lorentz transformation is, of course, 
anomaly-free for our vector-like model~\footnote{The
fact that the transformations $\delta\psi
=l(x)\gamma_{5}\psi$ and $\delta\psi
=c_{\mu}\gamma^{\mu}\gamma_{5}\psi$ are anomaly-free follows 
in superspace
from rewriting these transformations as 
$\delta\phi(x,\theta)=l(x)\bar{\theta}\gamma_{5}
(\partial/\partial\bar{\theta})
\phi(x,\theta)$ and 
$\delta\phi(x,\theta)=c_{\mu}(x)\bar{\theta}\gamma^{\mu}
\gamma_{5}(\partial/\partial\bar{\theta})\phi(x,\theta)$
respectively, and applying (3.17). The $d^{2}\theta$ integral 
projects out only the contributions 
$\bar{\theta}\gamma_{5}\theta$ and 
$\bar{\theta}\gamma^{\mu}\gamma_{5}\theta$, which vanish.} and 
its generator vanishes identically 
\begin{equation}
\bar{\psi}\gamma^{\mu}\gamma_{5}\psi\equiv 0.
\end{equation}
The  parameter 
$t^{\alpha}(x)$ generates the transformation 
$\delta F=\bar{t}\psi$, and it leads to the Ward identity 
containing the gamma-trace of the supercurrent.

The most interesting case are the transformations with $c_{\mu}(x)$. 
The parameter $c_{\mu}(x)$ generates the  transformations 
$\delta F=-\epsilon^{\mu\nu}c_{\mu}\partial_{\nu}\varphi$
and 
$\delta\psi=c_{\mu}\gamma^{\mu}\gamma_{5}\psi$, 
and these transformations yield the Ward identity
\begin{eqnarray}
c_{\mu}\tilde{\zeta}^{\mu}(x)
-\frac{1}{2}c_{\mu}\epsilon^{\mu\nu}
(\bar{\psi}\gamma_{\nu}\delslash\psi)
=c_{\mu}\zeta_{\mu}(x)
+c_{\mu}\frac{\hbar g}{2\pi}\epsilon^{\mu\nu}
\partial_{\nu}\varphi(x).
\end{eqnarray}
The last term in this Ward identity is an anomaly, and this 
anomaly constitutes the anomalous part in the central charge
current itself. This is an unusual point that may lead to 
confusion: the anomaly is proportional to the central charge current itself. 
In this respect it resembles neither the trace anomaly nor the chiral anomaly:
the trace anomaly contains a contraction of the current while the chiral 
anomaly contains a divergence of the current. In fact, it has been shown in~\cite{Rebhan:2002yw} 
that the central charge current is the anomaly in the divergence of the conformal central
charge current (which is explicitly $x$-dependent, just like the dilation current).
Actually, the anomaly comes only from the 
$\delta F$ variation and not from the $\delta\psi$ 
variation, see footnote 4. We shall later explicitly compute the 
anomaly from the $\delta F$ variation separately, see (4.19). 
In that case one finds the relation in (4.12) without the 
$\bar{\psi}\gamma_{\nu}\delslash\psi$ term
\begin{eqnarray}
\tilde{\zeta}^{\mu}(x)
=\zeta_{\mu}(x)
+\frac{\hbar g}{2\pi}\epsilon^{\mu\nu}\partial_{\nu}\varphi(x).
\end{eqnarray}
There is no 
contradiction: (4.12) should be used for correlation functions
with the variations $\delta F$ and $\delta\psi$, while (4.19) 
should be used for correlation functions with only $\delta F$.
Both relations are valid and different because the 
transformation rules are different.

{}From  (4.9) we obtain the relation
\begin{eqnarray}\label{4.100}
\tilde{T}_{\mu\nu}(x)-T_{\mu\nu}(x)&=&\frac{1}{2}
\eta_{\mu\nu}[-F^{2}-FU + \frac{1}{2}\bar{\psi}\delslash\psi+
g\varphi\bar{\psi}\psi]
\nonumber\\
&=&\eta_{\mu\nu}\frac{\hbar g}{4\pi}F
\end{eqnarray}
This relation, and others in (4.9), was obtained
by extracting contracted currents from the Ward identity, and 
by generalizing the contracted currents and the anomalies in the 
contracted currents to  the currents and the anomalies in the  
currents themselves. This does not determine the current 
completely. 
We now prove these uncontracted identities. We begin 
with~(\ref{4.100}) and 
claim the following relation
\begin{eqnarray}
\langle-F^{2}-FU + \frac{1}{2}\bar{\psi}\delslash\psi+
g\varphi\bar{\psi}\psi\rangle
&=&\langle-F(x)\frac{\delta S}{\delta F(x)}
-\frac{1}{2}\bar{\psi}(x)\frac{\delta S}{\delta\bar{\psi}(x)}
\rangle\nonumber\\
&=&\langle\frac{\hbar g}{2\pi}F\rangle.
\end{eqnarray}
Again we stress that this relation, and others to follow, holds
in Ward identities for correlation functions.
To prove this relation, consider 
\begin{eqnarray}
\delta\phi(x,\theta)&=&\frac{1}{2}w(x)\bar{\theta}
\frac{\partial}{\partial\bar{\theta}}\phi(x,\theta)
=\frac{1}{2}w(x)\bar{\theta}Q\phi(x,\theta)\nonumber\\
&=&\frac{1}{2}w(x)\bar{\theta}\psi(x)+w(x)\delta(\theta)F(x)
\end{eqnarray}
which is the ``R-symmetry'' in the present context and physically
corresponds to a Weyl transformation. Equating 
the variation of the action to that of the Jacobian in the 
path integral in the form of (4.4) gives the  
identity\footnote{The Noether current ${T^{N}}_{\mu\nu}(x)$
generated by the variation
$\delta\phi(x,\theta)=\xi^{\mu}(x)\partial_{\mu}\phi(x,\theta)$
is given by $\delta S=\int -(\partial^{\mu}\xi^{\nu})
{T^{N}}_{\mu\nu}$. It reads 
\begin{eqnarray}
{T^{N}}_{\mu\nu}(x)&=&\partial_{\mu}\varphi\partial_{\nu}\varphi
-\frac{1}{2}\eta_{\mu\nu}[(\partial^{\rho}\varphi)
(\partial_{\rho}\varphi)-F^{2}-2FU]\nonumber\\
&+&\frac{1}{4}\bar{\psi}[\gamma_{\mu}\partial_{\nu}+\gamma_{\nu}
\partial_{\mu}]\psi 
+\frac{1}{4}\bar{\psi}[\gamma_{\mu}\partial_{\nu}-\gamma_{\nu}
\partial_{\mu}]\psi 
-\frac{1}{2}\eta_{\mu\nu}
[\bar{\psi}\gamma^{\mu}\partial_{\mu}\psi
+2g\varphi \bar{\psi}\psi],\nonumber\\
&=&\tilde{T}_{\mu\nu}(x)-\eta_{\mu\nu}[-F^{2}-FU 
+ \frac{1}{2}\bar{\psi}\delslash\psi+g\varphi\bar{\psi}\psi]
+\frac{1}{4}\bar{\psi}[\gamma_{\mu}\partial_{\nu}-\gamma_{\nu}
\partial_{\mu}]\psi
\nonumber\\
&=&\tilde{T}_{\mu\nu}(x)-\eta_{\mu\nu}\frac{\hbar g}{2\pi}F
+\frac{1}{4}\bar{\psi}[\gamma_{\mu}\partial_{\nu}-\gamma_{\nu}
\partial_{\mu}]\psi
\nonumber
\end{eqnarray}
where we used the Weyl anomaly in (4.15). The last term is 
manifestly antisymmetric, and it can be written as
$\bar{\psi}[\gamma_{\mu}\partial_{\nu}
-\gamma_{\nu}\partial_{\mu}]\psi=-\epsilon_{\mu\nu}
\bar{\psi}\epsilon^{\rho\sigma}
\gamma_{\rho}\partial_{\sigma}\psi
=-\epsilon_{\mu\nu}
\bar{\psi}\gamma_{5}\gamma^{\rho}\partial_{\rho}\psi$.  
It yields the antisymmetric part of the stress tensor, and 
is proportional to the divergence of the Lorentz current.
 One may show that
$\partial^{\mu}{T^{N}}_{\mu\nu}(x)=
-\frac{\hbar g}{2\pi}\partial_{\nu}F(x)$ by evaluating  the 
Jacobian for $\delta\phi(x,\theta)
=\xi^{\mu}(x)\partial_{\mu}\phi(x,\theta)$. This Jacobian 
equals $-\xi^{\nu}(\frac{\hbar g}{2\pi}\partial_{\nu}F(x))$, 
and this implies 
$\partial^{\mu}\tilde{T}_{\mu\nu}(x)=0$. In our analysis, the 
conserved $\tilde{T}_{\mu\nu}$, and 
$T_{\mu\nu}$ which is not conserved but free of a trace anomaly, 
play a basic role. Note that both $\tilde{T}_{\mu\nu}$ and 
$T_{\mu\nu}$ are manifestly symmetric.
 } (4.15). This analysis fixes the magnitude of the Weyl 
anomaly. At the end of Section 5
we show that $T_{\mu}^{\mu}$ does not contribute to the 
trace anomaly, so the trace anomaly comes only from the 
conserved tensor.

The last relation in~(\ref{4.9}) to be proven is the one with 
$\tilde{J}^{\mu}$ and $j^{\mu}$.
We can show that 
\begin{eqnarray}
\langle \tilde{J}^{\mu}(x)-j^{\mu}(x)\rangle
&=&\langle\gamma^{\mu}\psi(x)(F(x)+U(\varphi))\rangle\nonumber\\
&=&\langle\gamma^{\mu}\psi(x)\frac{\delta S}{\delta F(x)}\rangle
=\langle-\frac{\hbar g}{2\pi}\gamma^{\mu}\psi(x)\rangle
\end{eqnarray}
by considering the variation
\begin{equation}\label{444}
\delta\phi(x,\theta)=\delta(\theta)\bar{\epsilon}_{\mu}(x)
\gamma^{\mu}\frac{\partial}{\partial\bar{\theta}}\phi(x,\theta)
\end{equation}
which agrees with~(\ref{4.9})\footnote{Clearly, the naive 
equation of motion $F(x)+U(\varphi)=0$ cannot be used in this 
derivation.}. 

Although we already obtained the relations between 
${\tilde \zeta}_{\mu}$ and $\zeta_{\mu}$ from~(4.7), we can 
apply the same techniques as used for ${\tilde T}_{\mu\nu}$ and 
$\tilde{J}^{\mu}$, and obtain then
\begin{eqnarray}
\langle\tilde{\zeta}_{\mu}(x)-\zeta_{\mu}(x)\rangle&=&
\langle-\epsilon_{\mu\nu}\partial^{\nu}\varphi(x)(F(x)+U)
\rangle\nonumber\\
&=&\langle-\epsilon_{\mu\nu}\partial^{\nu}\varphi(x)
\frac{\delta S}{\delta F(x)}\rangle
=\langle\frac{\hbar g}{2\pi}\epsilon_{\mu\nu}\partial^{\nu}
\varphi(x)\rangle
\end{eqnarray}
by considering the variation
\begin{equation}
\delta\phi(x,\theta)=-\delta(\theta)
v^{\mu}(x)\epsilon_{\mu\nu}\partial^{\nu}\phi(x,\theta)
\end{equation}
which is indeed consistent with~(\ref{4.9}). As we already 
discussed, the 
transformation~(4.19) generates only the central charge current
without inducing a variation of the fermion, and it generates the Ward identities with the 
central charge 
current itself, not its divergence. If one sets 
$v^{\mu}(x)=\partial^{\mu}v(x)$ in~(4.20), one generates the 
divergence of the central charge current but the procedure gives
no information for a topological current. In analogy with $U(1)$
gauge theory, we are considering the change of variable 
$A_{\mu}\rightarrow A_{\mu}+a_{\mu}$ instead of 
$A_{\mu}\rightarrow A_{\mu}+\partial_{\mu}a$ to generate the 
current.  

It is important to recognize that all operators 
appearing on the left-hand sides of the relations in (4.7)
have higher mass dimensions than those of the corresponding 
operators on the right-hand sides. For example, 
$\tilde{\zeta}^{\mu}(x)$ and $\zeta^{\mu}(x)$ are, respectively,
dimension $2$ and $1$ operators since the coupling constant 
$g$ carries a unit mass dimension. Similarly, 
$\gamma_{\mu}\tilde{J}^{\mu}$ and $\gamma_{\mu}j^{\mu}$ are, 
respectively, dimension $3/2$ and $1/2$ operators, though both of
$\tilde{J}^{\mu}$ and $j^{\mu}$ are dimension $3/2$
operators. Also, $\tilde{T}_{\mu}^{\ \mu}$ and $T_{\mu}^{\ \mu}$
are, respectively, dimension $2$ and $1$ operators, though
both of $\tilde{T}_{\mu\nu}$ and $T_{\mu\nu}$ are dimension
$2$ operators. In this sense all the composite operators on the 
right-hand sides of (4.7) are soft operators. This suggests that 
only the ``hard'' operators generate anomalies.  In the next 
section we prove this statement.

\section{Supersymmetry algebra of the quantum operators}

In the previous section we gave a direct derivation of the 
anomalies based on path integrals, but we already mentioned 
in the introduction that one can also obtain the anomalies from
the $\gamma_{\mu}\tilde{J}^{\mu}$ anomaly by making successive
susy transformations. In this section we implement this second 
approach. Since this involves commutators of currents, we 
convert the path integral relations 
into  operator relations by
following the BJL method.  

We begin by considering the variation
\begin{equation}
\delta\phi(x,\theta)=\bar{\epsilon}(x)Q\phi(x,\theta).
\end{equation}
The change of the action defines the Noether current
\begin{equation}
\delta S=\int d^{2}x(\partial_{\mu}\bar{\epsilon}(x))j^{\mu}(x)
\end{equation}
where
\begin{equation}
j^{\mu,\alpha}(x)=-\{[\delslash\varphi(x)+U(\varphi(x))]
\gamma^{\mu}\psi(x)\}^{\alpha}
\end{equation}
with $U(\varphi)=g(\varphi^{2}-v^{2}_{0})$. The Jacobian factor
for (5.1) gives the anomaly, and we obtain the 
identity
\begin{equation}
\partial_{\mu}j^{\mu}(x)=\frac{\hbar g}{2\pi}\delslash\psi(x)
\end{equation}
corresponding to the term proportional to $\delta(\theta)$
in (4.6). Thus the current $\tilde{J}^{\mu}$ defined 
in~(\ref{4.9}) is conserved
\begin{equation}\label{5.5}
\tilde{J}^{\mu}(x)\equiv j^{\mu}(x)
-\frac{\hbar g}{2\pi}\gamma^{\mu}\psi(x), \quad 
\partial_{\mu}\tilde{J}^{\mu}(x)=0.
\end{equation}
It contains the  contributions from the action and Jacobian,
appears in all the Ward identities, and this implies, as we shall see, that  the relations 
among various Green's functions obtained by global 
supersymmetry are not modified in form by  non-trivial 
Jacobians.

If one  considers the rigid conformal supersymmety transformation generated by the parameter 
\begin{equation}
\bar{\epsilon}(x)=\bar{a}(x)\xslash
\end{equation}
the action transforms as follows
\begin{equation}\label{act}
\delta S=-\int d^{2}x[(\partial_{\mu}\bar{a}(x))\xslash j^{\mu}(x)
+\bar{a}(x)\gamma_{\mu}j^{\mu}(x)],
\end{equation}
and one obtains the identity
\begin{equation}\label{cur}
\partial_{\mu}(\xslash \tilde{J}^{\mu}(x))
=\gamma_{\mu}j^{\mu}(x)-\frac{\hbar g}{\pi}
\psi(x)=\gamma_{\mu}\tilde{J}^{\mu}(x)
\end{equation}
The right hand side does not vanish in general, since 
the superconformal symmetry is explicitly broken in the 
present model.  If the action would have been
invariant under the transformation 
$\bar{\epsilon}(x)=\bar{a}\xslash$ with $x$-independent parameter
$\bar{a}$, the term $\gamma_{\mu}j^{\mu}$ would not have 
appeared 
in~(\ref{act}) and thus in the 
identity~(\ref{cur}). This shows that 
$\gamma_{\mu}j^{\mu}$ does not contain a superconformal anomaly.

We thus write the identity~(\ref{cur}) as
\begin{equation}
\partial_{\mu}(\xslash \tilde{J}^{\mu}(x))
=(\gamma_{\mu}j^{\mu}(x))_{exp}-\frac{\hbar g}{\pi}
\psi(x)=\gamma_{\mu}\tilde{J}^{\mu}(x)
\end{equation}
where $(\gamma_{\mu}j^{\mu}(x))_{exp}$ stands for the terms 
which explicitly break superconformal symmetry. Using 
$\gamma_{\mu}\gamma_{\nu}\gamma^{\mu}=0$ in strictly $d=2$ one
obtains
\begin{equation}
(\gamma_{\mu}\tilde{J}^{\mu}(x))_{exp}
=(\gamma_{\mu}j^{\mu})_{exp}=\gamma_{\mu}j^{\mu}
=-2U(\varphi)\psi(x)
\end{equation}
and 
\begin{eqnarray}
&&(\gamma_{\mu}\tilde{J}^{\mu}(x))_{anomaly}=-\frac{\hbar g}{\pi}
\psi(x),\nonumber\\
&&(\gamma_{\mu}j^{\mu}(x))_{anomaly}=0.
\end{eqnarray}  
We thus have two kinds of supersymmetry currents, 
$\tilde{J}^{\mu}$
and the Noether current $j^{\mu}(x)$. The current 
$\tilde{J}^{\mu}(x)$
is conserved but contains the superconformal anomaly, whereas
$j^{\mu}$ is not conserved but free from a  superconformal
anomaly.  Both contain explicit
conformal supersymmetry breaking terms.

The conserved current $\tilde{J}^{\mu}$ generates  
supersymmetry for all the components of $\phi(x,\theta)$. 
This can be shown by starting with 
\begin{equation}
\langle \phi(y,\theta)\rangle
=\int{\cal D}\mu\ \phi(y,\theta)\exp[iS]
\end{equation}
(recall that we suppress writing further fields $\phi_{1}(x_{1})
...\phi_{n}(x_{n})$)
and considering the change of variables (5.1).
One obtains then 
\begin{equation}
-i\partial_{\mu}\langle T^{\star}\tilde{J}^{\mu,\alpha}(x)
\phi(y,\theta)
\rangle
+\langle \delta_{susy}\phi(y,\theta)\rangle=0
\end{equation}
where $\delta_{susy}\phi(y,\theta)$ stands for the 
variation of $\phi(y,\theta)$
\begin{equation}
\delta_{susy}\phi(y,\theta)=\delta^{2}(x-y)
Q^{\alpha}\phi(x,\theta).
\end{equation}
We then apply the Bjorken-Johnson-Low (BJL) analysis to replace 
the $T^{\star}$
product by the $T$ product\footnote{The crucial point in the BJL 
analysis
in the present case is the treatment of the anomaly term
$\partial_{\mu}\langle T^{\star}\frac{\hbar g}{2\pi}
\gamma^{\mu}\psi(x)\phi(y,\theta)\rangle$. It 
should be replaced by 
$\partial_{\mu}\langle T\frac{\hbar g}{2\pi}
\gamma^{\mu}\psi(x)\phi(y,\theta)\rangle$ 
by noting that the $T$-product 
should satisfy the condition $\lim_{k_{0}\rightarrow\infty}
\int d^{2}k\exp[ik(x-y)]\langle T\frac{\hbar g}{2\pi}
\gamma^{\mu}\psi(x)\phi(y,\theta)\rangle$=0. If one would keep
 the derivative operator inside the $T$-product, this condition 
is spoiled. See the appendix for an account of the BJL 
prescription.
}
\begin{equation}
-i\partial_{\mu}\langle T \tilde{J}^{\mu,\alpha}(x)
\phi(y,\theta)\rangle+\langle \delta^{2}(x-y)
Q^{\alpha}\phi(x,\theta)\rangle=0
\end{equation}
and obtain in the limit $k_{0}\rightarrow\infty$ the equal time 
commutator (see appendix)
\begin{equation}
i[\tilde{J}^{0,\alpha}(x),\phi(y,\theta)]\delta(x^{0}-y^{0})
=\delta^{2}(x-y)[
\frac{\partial}{\partial\bar{\theta}_{\alpha}}
-(\gamma^{\mu}\theta)^{\alpha}\partial_{\mu}]\phi(x,\theta).
\end{equation}
This proves the conservation equation
$\langle T\partial_{\mu}\tilde{J}^{\mu,\alpha}(x)
\phi(y,\theta)\rangle
=0$, and shows that $\tilde{J}^{0,\alpha}$ generates  
ordinary supersymmetry transformations..

We next specify the energy-momentum tensor $T_{\mu}^{\ \nu}$
and the central charge current $\zeta_{\mu}(x)$.
The charge $\tilde{Q}^{\alpha}$ 
\begin{equation}
\tilde{Q}^{\alpha}=\int dx \tilde{J}^{0,\alpha}(x)
\end{equation}
generates supersymmetry, as we have shown above, and thus we 
have
\begin{eqnarray}\label{ppp}
i[\bar{\epsilon}\tilde{Q}, j^{\nu,\beta}(y)]
&=&-2T^{\ \nu}_{\mu}(y)
(\gamma^{\mu})^{\beta}_{\ \alpha}\epsilon^{\alpha}
-2\zeta^{\nu}(y)
(\gamma_{5})^{\beta}_{\ \alpha}\epsilon^{\alpha} 
-(F+U)(\delslash\varphi(y)\gamma^{\nu}\epsilon)^{\beta}
\nonumber\\
&=&-2T^{\ \nu}_{\mu}(y)
(\gamma^{\mu})^{\beta}_{\ \alpha}\epsilon^{\alpha}
-2\zeta^{\nu}(y)
(\gamma_{5})^{\beta}_{\ \alpha}\epsilon^{\alpha} 
+\frac{\hbar g}{2\pi}(\delslash\varphi(y)\gamma^{\nu}
\epsilon)^{\beta}
\end{eqnarray}
where we used (4.19). In the path integral framework, this 
relation is derived by starting with  
$\langle j^{\nu}(y)\rangle =\int {\cal D}\phi j^{\nu}(y)
e^{iS}$
and considering the change of variables corresponding to 
(local) supersymmetry
\begin{eqnarray}
i\partial_{\mu}\langle T^{\star}\tilde{J}^{\mu}(x)j^{\nu}(y)
\rangle
=\delta(x-y)\langle \delta_{susy}j^{\nu}(y)\rangle.
\end{eqnarray}
The local variations of the action and the measure give 
together the left-hand side, just as in (5.13). 
The BJL analysis then gives rise to the commutator.
The operators appearing here are given 
by\footnote{By noting the 
completeness of $(1,\gamma_{5},\gamma^{\mu})$, 
the supersymmetry variation of the current $j^{\nu}(y)$
is expanded as 
\begin{eqnarray}
\delta_{\epsilon}j^{\nu,\beta}(y)=-2T^{\ \nu}_{\mu}(y)
(\gamma^{\mu})^{\beta}_{\ \alpha}\epsilon^{\alpha}
-2\zeta^{\nu}(y)
(\gamma_{5})^{\beta}_{\ \alpha}\epsilon^{\alpha}
-2v^{\nu}(y)\epsilon^{\beta}\nonumber
\end{eqnarray}
By multiplying this relation by $\bar{\epsilon}\gamma_{\rho}$,
$\bar{\epsilon}\gamma_{5}$ and $\bar{\epsilon}$,respectively,
 we can project out the 3 components above by noting 
$\bar{\epsilon}\gamma^{\mu}\epsilon=\bar{\epsilon}\gamma_{5}
\epsilon=0$. The vector 
component $v^{\nu}$ is shown to vanish on-shell by using  
safe (i.e., anomaly-free) relations such as 
$\bar{\psi}(x)\gamma^{\mu}\frac{\delta S}{\delta\bar{\psi}(x)}
=\bar{\psi}(x)\gamma_{5}\gamma^{\mu}
\frac{\delta S}{\delta\bar{\psi}(x)}=0$, except for the 
term explicitly written in~(\ref{ppp}).}
\begin{eqnarray}
\zeta_{\mu}(x)&=&\epsilon_{\mu\nu}\partial^{\nu}\varphi(x)
U(\varphi),\ \ \ \epsilon_{01}=-1,\nonumber\\
T_{\mu\nu}(x)&=&\partial_{\mu}\varphi\partial_{\nu}\varphi
-\frac{1}{2}\eta_{\mu\nu}[(\partial^{\rho}\varphi)
(\partial_{\rho}\varphi)-UF]\nonumber\\
&+&\frac{1}{4}\bar{\psi}[\gamma_{\mu}\partial_{\nu}+\gamma_{\nu}
\partial_{\mu}]\psi
-\frac{1}{4}\eta_{\mu\nu}
[\bar{\psi}\gamma^{\mu}\partial_{\mu}\psi
+2g\varphi\bar{\psi}\psi]
\end{eqnarray}
which agree with the operators in (4.9).

We thus obtain
\begin{eqnarray}
i[\bar{\epsilon}\tilde{Q}, \gamma_{\nu}j^{\nu}_{\beta}(y)]
&=&-2T^{\ \mu}_{\mu}(y)\epsilon
-2\gamma_{\nu}\zeta^{\nu}(y)\gamma_{5}\epsilon\nonumber\\
&=&-2(FU-g\varphi\bar{\psi}\psi)\epsilon
-2(\gamma_{\nu}\epsilon^{\nu\mu}\partial_{\mu}\varphi(x)U)
\gamma_{5}\epsilon
\end{eqnarray}
Since the operator $\gamma_{\mu}j^{\mu}$ does not contain
an anomaly, the operators on the right-hand side are expected  
not to contain superconformal anomalies either
\begin{equation}
T_{\mu}^{\ \mu}(x)_{anomaly}=0, \ \ \ \ \ 
\gamma_{\mu}\zeta^{\mu}(x)_{anomaly}=0.
\end{equation}
The absence of an anomaly in $T_{\mu}^{\ \mu}$ is consistent 
with the analysis in the previous section. 
If $\gamma_{\mu}j^{\mu}=0$, these operators also vanish.
The operator $T_{\mu}^{\ \nu}$ is not conserved
\begin{equation}
\partial_{\nu}T_{\mu}^{\ \nu}(x)\neq 0
\end{equation}
due to the effects of the anomaly, in accord with the 
non-conservation of the Noether current $j^{\nu}$. We emphasize 
that 
$T_{\mu}^{\ \mu}(x)$ and $T_{\nu}^{\ \mu}(x)$ should be 
clearly distinguished, and in fact one cannot reproduce
$T_{\nu}^{\ \mu}(x)$ from the knowledge of 
$T_{\mu}^{\ \mu}(x)$ alone. For this reason, 
it is possible that $T_{\mu}^{\ \mu}(x)$ does not contain 
anomaly while $T_{\nu}^{\ \mu}(x)$ does contain anomalies.
 Either the conservation 
condition or the trace is influenced by the anomaly. For 
$T_{\mu}^{\ \mu}(x)$, we have no conservation condition.

We next examine the supersymmetry algebra for 
the conserved supercurrent
\begin{equation}
i[\bar{\epsilon}\tilde{Q}, \tilde{J}^{\nu\ \beta}(y)]
=-2\tilde{T}^{\ \nu}_{\mu}(y)
(\gamma^{\mu})^{\beta}_{\ \alpha}\epsilon^{\alpha}
-2\tilde{\zeta}^{\nu}(y)
(\gamma_{5})^{\beta}_{\ \alpha}\epsilon^{\alpha}
\end{equation}
where 
\begin{eqnarray}
\partial_{\nu}\tilde{T}^{\ \nu}_{\mu}(y)=0,\ \ \ \ \ 
\partial_{\nu}\tilde{\zeta}^{\nu}(y)=0
\end{eqnarray}
because of the conservation $\partial_{\nu}\tilde{J}^{\nu}=0$.
Again this relation is obtained by making a local supersymmetry
transformation as in (5.13) and (5.19).
If one uses the expression
\begin{equation}
\tilde{J}^{\mu}(x)=-[\delslash\varphi(x)-F]
\gamma^{\mu}\psi(x)
\end{equation}
in the path integral
\begin{equation}
\int{\cal D}\phi \tilde{J}^{\nu}(y)
\exp[iS]
\end{equation}
and considers the change of variables corresponding
to a supersymmetry transformation, one obtains
\begin{eqnarray}
i\partial_{\mu}\langle T^{\star}\tilde{J}^{\mu}(x)
\tilde{J}^{\nu}(y)\rangle
=\delta(x-y)\langle \delta_{susy}\tilde{J}^{\nu}(y)\rangle.
\nonumber
\end{eqnarray}
(To be precise, the current $\tilde{J}^{\mu}(x)$ on the 
left-hand side is defined by (5.5).)
The operator commutation relation $i[\bar{\epsilon}\tilde{Q},
\tilde{J}^{\nu\ \beta}(y)]$ is then
obtained by the BJL analysis. In this way
one obtains the expressions for $\tilde{T}_{\mu\nu}(y)$ and 
$\tilde{\zeta}^{\nu}(y)$ in (4.9). 

Instead, if one uses
the form of $\tilde{J}^{\mu}=j^{\mu}-
\frac{\hbar g}{2\pi}\gamma^{\mu}\psi$ and~(\ref{ppp}) and 
repeats the 
same procedure, one obtains (in this case, we have two identical
 $\tilde{J}^{\mu}$)
\begin{eqnarray}
\tilde{T}_{\mu\nu}(y)&=&T_{\mu\nu}(y)
+\eta_{\mu\nu}\frac{\hbar g}{4\pi}F,\nonumber\\
\tilde{\zeta}^{\nu}(y)&=&\zeta^{\nu}(y)+\frac{\hbar g}{2\pi}
\epsilon^{\nu\mu}\partial_{\mu}\varphi(y).
\end{eqnarray}
Note that the extra term in~(\ref{ppp}) combines with a part of the 
variation
of $-\frac{\hbar g}{2\pi}\gamma^{\mu}\psi$ to give rise to 
the central charge anomaly in the algebra for conserved current.

We next note that 
\begin{eqnarray}
i[\bar{\epsilon}\tilde{Q},\gamma_{\nu}\tilde{J}^{\nu}(y)]
&=&-2\tilde{T}^{\ \mu}_{\mu}(y)\epsilon
-2\gamma_{\nu}\tilde{\zeta}^{\nu}(y)
\gamma_{5}\epsilon
\nonumber\\
&=&-2[T^{\ \mu}_{\mu}(y)+\frac{\hbar g}{2\pi}F(y)]
\epsilon\nonumber\\
&&-2[\gamma_{\nu}\zeta^{\nu}(y)
\gamma_{5}\epsilon
+\frac{\hbar g}{2\pi}
\partial_{\mu}\varphi(y)\gamma^{\mu}\epsilon]
\end{eqnarray}
where the second expression  is obtained by the super
transformation generated by $\tilde{Q}$ of the expression 
\begin{equation}
\gamma_{\nu}\tilde{J}^{\nu}(y)=\gamma_{\nu}j^{\nu}(y)
-\frac{\hbar g}
{\pi}\psi(y)=-2U(\varphi)\psi(y)-\frac{\hbar g}
{\pi}\psi(y).
\end{equation}
We thus have
\begin{eqnarray}
&&\tilde{T}^{\ \mu}_{\mu}(x)=T_{\mu}^{\ \mu}(x)
+\frac{\hbar g}{2\pi}F(x),\nonumber\\
&&\gamma_{\nu}\tilde{\zeta}^{\nu}(x)=
\gamma_{\mu}\zeta^{\mu}(x)
+\frac{\hbar g}{2\pi}
\partial_{\mu}\varphi(x)\gamma^{\mu}\gamma_{5}.
\end{eqnarray}

We would like to explain that the separation of anomaly
terms from the explicit (soft) symmetry breaking terms makes
sense already in the trivial vacuum and to order $\hbar g$.
We start with the renormalization of the vacuum value 
by a tadpole contribution. We adopt the following minimal
renormalization convention~\cite{Rebhan:1997iv,Rebhan:2002uk}
\begin{equation}
U=g(\varphi^{2}-v^{2}_{0})=g(\varphi^{2}-v^{2})
-g\delta v^{2}=U_{ren}+\Delta U
\end{equation}
at the one-loop level accuracy in the trivial vacuum. 
We then analyze the explicit breaking term in the central charge
current
\begin{eqnarray}
\zeta_{\mu}(x)&=&\epsilon_{\mu\nu}\partial^{\nu}\varphi(x)
U(\varphi)\nonumber\\
&=&g\epsilon_{\mu\nu}\partial^{\nu}\varphi(x)
(\varphi^{2}-v^{2}_{0}).
\end{eqnarray}

Our regularization of the Jacobian by the heat kernel in~(2.19) and~(2.30) corresponds
to the following regularization of the propagator 
\begin{equation}\label{tup}
\langle T^{\star}\eta(x,\theta)\eta(y,\theta^{\prime})\rangle
=\frac{i}{\Gamma}\exp[\frac{\Gamma^{2}}{M^{2}}]\delta(x-y)
\delta(\theta-\theta^{\prime})
\end{equation}
The factor $\exp[\Gamma^{2}/M^{2}]$ gives an 
exponential cut-off whose leading term is 
$\exp[-\frac{1}{4}k^{2}/M^{2}]$. Since there is a coupling 
constant $g$ in front of
$\zeta_{\mu}$, only the tadpole diagrams are relevant to 
order $\hbar g$. The tadpole $\langle \varphi(x)\varphi(x)
\rangle$
is absorbed into the renormalization of the vacuum value. The 
remaining tadpole has the form 
\begin{equation}
\langle \partial_{\mu}\varphi(x)\varphi(x)\rangle=0
\end{equation}
since, to this order, the propagator becomes the free one and 
our regulator becomes $\exp[-\frac{1}{4}p^{2}/M^{2}]$.
Lorentz symmetry thus gives a vanishing result:
the operator $\zeta_{\mu}$ does not induce an anomaly
term\footnote{This conclusion also holds in some other 
regularization schemes such as the dimensisonal 
regularization and the dimensional 
reduction~\cite{Rebhan:2002yw}. On the 
other hand,
using point splitting, the operator 
$\int\delta(x-y)(\partial_{\sigma}\varphi(x)U(y)
+\partial_{\sigma}\varphi(y)U(x))dx$ does contain the 
anomaly~\cite{GLN}. Setting naively $x=y$ in their expressions,
one loses the anomaly and obtains $\zeta_{\sigma}(y)$.
 } to order $\hbar g$. 

On the other hand, the  ``hard'' operator 
$\tilde{\zeta}_{\mu}(x)
=-\epsilon_{\mu\nu}\partial^{\nu}\varphi(x)F(x)$, which is of 
zeroth order in $g$, produces an anomaly to order 
$\hbar g$, in addition to the soft breaking term $\zeta_{\mu}$. 
In this case we need to use the full expression of 
the propagator in~(\ref{tup}) and this gives rise to the one 
loop 
anomaly as was shown by a superfield calculation in Section 4.  

We finally examine the soft breaking term of conformal symmetry
\begin{equation}
T_{\mu}^{\ \mu}(x)=FU-g\varphi\bar{\psi}\psi
=gF(\varphi^{2}-v^{2}_{0})-g\varphi\bar{\psi}\psi
\end{equation}
This operator  does not generate a  trace anomaly in the 
trivial vacuum to order $\hbar g$, since this combination of
 composite operators appears inside the action, where the 
renormalization prescription of these composite operators is 
precisely specified. One can also confirm this statement by an 
analysis of one-loop tadpole diagrams; the tadpole
$\langle\varphi(x)\varphi(x)\rangle$ is absorbed by the 
renormalization of the vacuum value, and 
$2\langle\varphi(x)F(x)\rangle$
and $\langle\bar{\psi}(x)\psi(x)\rangle$ precisely cancel each
other\footnote{The same conclusions are reached for the kink 
vacuum: the operator $T_{\mu}^{\ \mu}$ does not contribute
 to the anomaly. }.
We prove this statement by a manifestly 
supersymmetric calculation
\begin{eqnarray}
T_{\mu}^{\ \mu}(x)&=&
gF(x)(\varphi^{2}(x)-v^{2}_{0})-g\varphi(x)\bar{\psi}(x)\psi(x)
\nonumber\\
&=&\int d^{2}\theta g[\frac{1}{3}\phi^{3}(x,\theta)
-v^{2}_{0}\phi(x,\theta)]\nonumber
\end{eqnarray}
which gives rise to the following $\hbar g$ contribution in the 
trivial vacuum
\begin{eqnarray}
\delta T_{\mu}^{\ \mu}(x)&=&
\int d^{2}\theta g[\phi(x,\theta)\langle T^{\star}\phi(x,\theta)
\phi(x,\theta)\rangle-\hbar\delta v^{2}\phi(x,\theta)]\nonumber\\
&=&\int d^{2}\theta g\phi(x,\theta)
[\langle T^{\star}\eta(x)\eta(x)\rangle-\hbar\delta v^{2}]=0.
\end{eqnarray}
In the first line we encounter both of 
a $\langle\bar{\psi}\psi\rangle$ and a $\langle\varphi F\rangle$
terms, and in the second line we used (2.22) for 
$\theta_{1}=\theta_{2}$. Hence $2\langle\varphi(x)F(x)\rangle$
is indeed equal to $\langle\bar{\psi}(x)\psi(x)\rangle$. 

When one analyzes the kink vacuum, one obtains non-anomalous
one-loop corrections from these soft operators (though these
corrections vanish after spatial integration).

\section{The BPS bound}

We have identified the quantum operators in 
the supersymmetry algebra and deduced the presence of  
 superconformal anomalies from a 
manifestly supersymmetric superfield calculation. We next 
 examine how the anomalies modify the BPS bound 
for a kink solution. 

The algebra of supersymmetry charges follows from (5.24) and
reads
\begin{eqnarray}
i\{\tilde{Q}^{\alpha},\tilde{Q}^{\beta}\}
&=&-2(\gamma^{\mu})^{\alpha\beta}\tilde{P}_{\mu}
-2\tilde{Z}(\gamma_{5})^{\alpha\beta}
\end{eqnarray}
where we defined 
\begin{eqnarray}
&&\tilde{P}_{\mu}=\int dx \tilde{T}_{0\mu}(x), 
\ \ \ \tilde{H}=\tilde{P}_{0} ,\nonumber\\
&&\tilde{Z}=\int dx \tilde{\zeta}_{0}(x)
=-\int dx\tilde{\zeta}^{0}(x)
\end{eqnarray} 
We can now evaluate the modified energy and central charge in 
the vacuum  of a time independent kink solution 
$\varphi_{K}(x)$, whose center is located at the origin, (to the 
accuracy of $O(\hbar)$) 
\begin{equation}
\langle0| \varphi(x)|0\rangle=
\varphi_{K}(x)+\hbar\varphi_{1}(x)\equiv v\tanh(gvx)
+\hbar\varphi_{1}(x)
\end{equation}
where $\hbar\varphi_{1}(x)$ is a quantum correction to the 
kink solution~\cite{Shifman:1998zy, GLN}. The results are 
\begin{eqnarray}
\langle0|\tilde{H}|0\rangle&=&-\langle0|
\int dx \tilde{T}_{0}^{\ 0}(x)|0
\rangle=-\langle0|\int dx[\tilde{T}_{0}^{\ 0}(x)+
\tilde{T}_{1}^{\ 1}(x)]|0\rangle\nonumber\\
&=&-\langle0|\int dx[T_{\mu}^{\ \mu}(x)
+\frac{\hbar g}{2\pi}F(x)]|0\rangle\nonumber\\
&=&-\langle0|\int dx[(FU-g\varphi\bar{\psi}\psi)
+\frac{\hbar g}{2\pi}F(x)]|0\rangle\nonumber\\ 
&=&M-\frac{\hbar m}{2\pi}
\end{eqnarray}
where $M$ is the classical mass of the kink solution, 
 and $m=2gv$ 
stands for the fermion mass at spatial infinity. Furthermore 
\begin{eqnarray}
\langle0|\tilde{Z}|0\rangle&=&
-\langle0|\int dx \tilde{\zeta}^{0}|0\rangle
=-\langle0|\frac{1}{2}tr\int dx\gamma^{0}
[\gamma_{0}\tilde{\zeta}^{0}+\gamma_{1}\tilde{\zeta}^{1}]
|0\rangle
\nonumber\\
&=&-\langle0|\frac{1}{2}tr\int dx\gamma^{0}[\gamma_{\mu}
\zeta^{\mu}(x)
+\frac{\hbar g}{2\pi}
\partial_{\mu}\varphi(x)\gamma^{\mu}\gamma_{5}]|0\rangle
\nonumber\\
&=&-\langle0|\frac{1}{2}tr\int dx\gamma^{0}
[(\gamma_{\nu}\epsilon^{\nu\mu}
\partial_{\mu}\varphi(x)U)+\frac{\hbar g}{2\pi}
\partial_{\mu}\varphi(x)\gamma^{\mu}\gamma_{5}]|0\rangle
\nonumber\\
&=&-\langle0|\int dx
\{g\partial_{1}\varphi(x)[\varphi^{2}(x)-v^{2}_{0}]
+\frac{\hbar g}{2\pi}\partial_{1}\varphi(x)\}|0\rangle
\nonumber\\
&=&M-\frac{\hbar m}{2\pi}.
\end{eqnarray}    
Here we used (6.10) below in the last step. 

In the first step
of both of (6.4) and (6.5) we used  
\begin{equation}
\langle0| \int dx\tilde{\zeta}^{1}(x)|0\rangle=0, \ \ \ \
\langle0| \int dx\tilde{T}_{1}^{\ 1}|0\rangle=0
\end{equation}
which are the consequences of the conservation condition of 
$\tilde{\zeta}^{\mu}(x)$ and $\tilde{T}_{\mu}^{\ \nu}$
\begin{eqnarray}
&&\langle0|[\partial_{0}\tilde{\zeta}^{0}(x)
+\partial_{1}\tilde{\zeta}^{1}(x)]|0\rangle
=\partial_{1}\langle0|\tilde{\zeta}^{1}(x)|0\rangle=0,
\nonumber\\
&&\langle0|[\partial_{0}\tilde{T}^{\ 0}_{1}(x)
+\partial_{1}\tilde{T}^{\ 1}_{1}(x)]|0\rangle
=\partial_{1}\langle0|\tilde{T}^{\ 1}_{1}(x)|0\rangle=0.
\end{eqnarray}
Namely,$\langle0|\tilde{\zeta}^{1}(x)|0\rangle$ and 
 $\langle0|\tilde{T}^{\ 1}_{1}(x)|0\rangle$ are 
independent of $x$, which may be fixed at the values at 
spatial infinity $\langle0|\tilde{\zeta}^{1}(\infty)|0\rangle=0$ 
and $\langle0|\tilde{T}^{\ 1}_{1}(\infty)|0\rangle=0$ (or 
$\langle0|\tilde{\zeta}^{1}(0)|0\rangle=0$ due to parity 
invariance of the kink vacuum). We note 
that the particle spectrum at spatial infinity has a 
well-defined mass gap, and in the 
asymptotic region of the kink vacuum there is no flow of central
 charge, namely, 
$\langle0|\tilde{\zeta}^{1}(\infty)|0\rangle=0$,
and no pressure, namely, 
$\langle0|\tilde{T}^{\ 1}_{1}(\infty)|0\rangle=0$.

We now explain that the BPS bound is generally satisfied 
for our kink solution and for our formulas of 
$\langle0|\tilde{H}|0\rangle$ and $\langle0|\tilde{Z}|0\rangle$
in (6.4) and (6.5). We start with the analysis of
\begin{eqnarray}
\langle0|
Q^{\alpha}\frac{\hbar g}{2\pi}\phi(x,\theta)|0\rangle
=\frac{\hbar g}{2\pi}\langle0|\theta^{\alpha} F(x)
-(\gamma^{1}\theta)^{\alpha}
\partial_{1}\varphi(x)|0\rangle
\end{eqnarray}
where $Q^{\alpha}$ is the differential operator in (2.8). We 
also use the 
fact that $\partial_{0}\langle0|O(t,x)|0\rangle=0$
for a general operator $O(t,x)$ and that the expectation values
of the fermionic components vanish in (6.8). When one chooses 
$\alpha=+$ in the above relation, one obtains 
\begin{eqnarray}
\langle0|
Q^{+}\frac{\hbar g}{2\pi}\phi(x,\theta)|0\rangle
=\theta^{+}\frac{\hbar g}{2\pi}\langle0|[F(x)-
\partial_{1}\varphi(x)]|0\rangle=0
\end{eqnarray}
since $i[\tilde{Q}^{+}, \phi]=Q^{+}\phi$ according to (5.16),
while $\tilde{Q}^{+}$ preserves the kink 
vacuum~\footnote{The classical kink solution which is 
specified by $\varphi_{K}(x)$ and $F_{K}(x)$, 
$F_{K}(x)=\partial_{1}\varphi_{K}(x)$, preserves 
$\tilde{Q}^{+}$ symmetry and $\tilde{Q}^{+}$ has no ordinary- 
supersymmetry anomaly ($\tilde{J}_{\mu}$ is conserved), while
the action is fully 
supersymmetric. Thus the $\tilde{Q}^{+}$ symmetry is 
preserved  to all orders in quantum corrections.}.
 This  identity shows that the anomalous 
contributions to 
$\langle0|\tilde{H}|0\rangle$ in (6.4) and 
$\langle0|\tilde{Z}|0\rangle$ in (6.5)
are identical.

Similarly, one can show that the nonanomalous contributions
to $\langle0|\tilde{H}|0\rangle$ and 
$\langle0|\tilde{Z}|0\rangle$ are equal. We use the identity 
\begin{eqnarray}
&&\langle0|
Q^{+}[\frac{g}{3}\phi^{3}(x,\theta)-gv^{2}_{0}\phi(x,\theta)]
|0\rangle\nonumber\\
&&=\theta^{+}g\langle0|\{[F(x)(\varphi(x)^{2}-v^{2}_{0})
-\varphi(x)\bar{\psi}\psi(x)]-
\partial_{1}\varphi(x)[\varphi^{2}(x)-v^{2}_{0}]\}|0\rangle
\nonumber\\
&&=0
\end{eqnarray}   
which shows that the contributions from explicit superconformal
symmetry breaking terms to both of 
$\langle0|\tilde{H}|0\rangle$ and $\langle0|\tilde{Z}|0\rangle$
(and also to their densities) are identical in the kink vacuum 
which preserves $Q^{+}$ symmetry. This statement holds to all 
orders in quantum 
corrections as far as the kink vacuum preserves $Q^{+}$ 
symmetry.  

We next prove that higher order nonanomalous corrections to the 
mass $M$ of the kink solution are absent.  
We define the {\em exact} solution which includes higher order
quantum corrections to the classical kink solution
\begin{equation}
\langle0|\varphi(x)|0\rangle=\varphi_{c}(x)
\end{equation}
with  the boundary condition (see (6.3) to the order $O(\hbar)$)
\begin{equation}
\varphi_{c}(\pm\infty)=\varphi_{K}(\pm\infty)
\end{equation}
and set
\begin{equation}
\varphi(x)=\varphi_{c}(x)+\eta(x).
\end{equation}
We  evaluate the central charge (6.5) arising from the explicit 
superconformal symmetry breaking
\begin{eqnarray}
&&-g\int dx \langle0|
\partial_{1}\varphi(x)[\varphi^{2}(x)-v^{2}_{0}]|0\rangle
\nonumber\\
&&=-g\int dx\partial_{1}\varphi_{c}(x)[\varphi_{c}^{2}(x)-v^{2}]
-g\int dx\partial_{1}\varphi_{c}(x)\langle0|[\eta(x)\eta(x)
-\delta v^{2}]|0\rangle\nonumber\\
&&-2g\int dx\varphi_{c}(x)\langle0|[\partial_{1}\eta(x)\eta(x)]
|0\rangle
-g\int dx\langle0|\partial_{1}\eta(x)\eta(x)\eta(x)|0\rangle
\end{eqnarray}
where $\delta v^{2}$ in this expression includes all the 
higher order quantum corrections.
The first term in (6.14) gives the classical kink mass
\begin{equation}
-g\int dx\partial_{1}\varphi_{c}(x)[\varphi_{c}^{2}(x)-v^{2}]
=\frac{4}{3}gv^{3}=M
\end{equation}
The second and the third terms in (6.14) together give
\begin{eqnarray}
&&-g\int dx\partial_{1}\varphi_{c}(x)\langle0|[\eta(x)\eta(x)
-\delta v^{2}]|0\rangle
-2g\int dx\varphi_{c}(x)\langle0|[\partial_{1}\eta(x)\eta(x)]
|0\rangle
\nonumber\\
&&=-g\int dx\partial_{1}\{\varphi_{c}(x)\langle0|[\eta(x)\eta(x)
-\delta v^{2}]|0\rangle\}=0
\end{eqnarray}
since (nonperturbatively)
\begin{equation}
\langle0|[\eta(x)\eta(x)
-\delta v^{2}]|0\rangle|_{x=\pm\infty}=0
\end{equation}
because of the renormalization condition in the trivial vacuum.
The last term in (6.14) gives
\begin{eqnarray}
-g\int dx\langle0|\partial_{1}\eta(x)\eta(x)\eta(x)|0\rangle
=-g\frac{1}{3}\int dx\partial_{1}
\langle0|\eta(x)\eta(x)\eta(x)|0\rangle=0
\end{eqnarray}
by noting 
\begin{equation}
g\langle0|\eta(x)\eta(x)\eta(x)|0\rangle|_{x=\infty}
=g\langle0|\eta(x)\eta(x)\eta(x)|0\rangle|_{x=-\infty}.
\end{equation}
This last relation arises from the fact that 
we have  in the trivial vacuum $|v\rangle$
\begin{equation}
g\langle v|\eta(x)\eta(x)\eta(x)|v\rangle|_{x=\infty}
=g\langle v|\eta(x)\eta(x)\eta(x)|v\rangle|_{x=-\infty}
\end{equation}
because of the translational invariance. The theory in the 
trivial vacuum  is invariant under the replacement
\begin{eqnarray}
v\rightarrow -v,\ \ \ \ g\rightarrow -g, \ \ \ \eta(x)
\rightarrow-\eta(x), \ \ \ F(x)\rightarrow -F(x)
\end{eqnarray}
and thus 
\begin{equation}
g\langle v|\eta(x)\eta(x)\eta(x)|v\rangle|_{x=-\infty}
=g\langle-v|\eta(x)\eta(x)\eta(x)|-v\rangle|_{x=-\infty}
\end{equation}
which leads to (6.19).
Note that the operator 
\begin{equation}
\eta(x)\eta(x)\eta(x)-3\delta v^{2}\eta(x)
\end{equation}
is finite since only the tadpole of $\eta$ is divergent even in 
the kink vacuum. Thus\\ 
$\langle0|\eta(x)\eta(x)\eta(x)|0\rangle$
is finite since $\langle0|\eta(x)|0\rangle=0$.

To the acuracy of $O(\hbar)$, the energy and central charge
densities receive the same amount of correction
\begin{equation}
-g\partial_{1}\{\varphi_{K}(x)\langle0|[\eta(x)\eta(x)
-\hbar\delta v^{2}]|0\rangle\}
\end{equation}
which vanishes after spatial integration. The explicit 
evaluation of $\langle0|[\eta(x)\eta(x)
-\hbar\delta v^{2}]|0\rangle$ has been performed 
in~\cite{Shifman:1998zy, GLN} and 
reads\footnote{The term with $-1$ is due to the counter term 
for the vacuum value renormalization, and 
the bound state has $\omega_{B}=\frac{1}{2}\sqrt{3}m$ and 
$\varphi_{B}(x)=\sqrt{\frac{3m}{4}}
\frac{\sinh(mx/2)}{\cosh^{2}(mx/2)}$. The continuum 
solutions read\\
$\varphi(k,x)=e^{ikx}[-3\tanh^{2}\frac{mx}{2}+1+
4(\frac{k}{m})^{2}+6i\frac{k}{m}\tanh \frac{mx}{2}]/N$
with $\omega=\sqrt{k^{2}+m^{2}}$ and $N^{2}=
16\frac{\omega^{2}}{m^{2}}(\frac{\omega^{2}}{m^{2}}-
\frac{\omega^{2}_{B}}{m^{2}})$. }
\begin{eqnarray}
\langle0|[\eta(x)\eta(x)-\hbar\delta v^{2}]|0\rangle
&=&\int\frac{dk}{2\pi}
\frac{1}{2\omega}
(|\varphi(k,x)|^{2}-1)+\frac{1}{\omega_{B}}\varphi^{2}_{B}(x)
\nonumber\\
&=&-\frac{3}{8\pi}\frac{1}{\cosh^{4}\frac{mx}{2}}+
\frac{1}{4\sqrt{3}}\frac{\sinh^{2}
\frac{mx}{2}}{\cosh^{4}\frac{mx}{2}}
\end{eqnarray}
which obviously satisfies the boundary conditions in (6.17).

The anomaly term gives rise to 
\begin{eqnarray}
-\frac{\hbar g}{2\pi}\int dx\partial_{1}
\langle|\varphi(x)|0\rangle
=-\frac{\hbar g}{2\pi}\int dx\partial_{1}\varphi_{c}(x)
=-\frac{2\hbar gv}{2\pi}=-\frac{\hbar m}{2\pi}.
\end{eqnarray}
We note that the total central charge can be written as (to the 
accuracy of $O(\hbar)$) 
\begin{equation}
M-\frac{\hbar m}{2\pi}
=\frac{4}{3}g(v^{2}-\frac{\hbar}{2\pi})^{3/2}
\end{equation}
which shows that the superconformal anomalies could be 
effectively represented
as a well-defined finite shift of the renormalized vacuum 
value.

We thus find that  the BPS bound is maintained as a result of
the uniform shifts in energy and central charge in the kink 
vacuum, to the accuracy of the present approximation.
Our analysis of the ground state with a kink solution   
is consistent  with previous  analyses 
on the basis of various other regularization 
methods~\cite{Nastase:1998sy, Shifman:1998zy, Graham:1998qq,
Rebhan:2002yw, Bordag:2002dg, Rebhan:2002uk, GLN}.
\\

It is instructive to consider the problem from the point of view
of a quantum deformation of the supersymmetry algebra by 
anomalies. This deformation appears if one expresses the 
right-hand side of the supersymmetry algebra in terms of 
the (non-conserved) operators\footnote{The use of 
non-conserved quantities in the presence of background metric
is common in conformal field theory,
and the anomalous conservation of $T_{\mu}^{\ \nu}$ there gives 
rise 
to the central charge in the Virasoro algebra~\cite{flnrvn}.} 
$T_{\mu}^{\ \nu}$ and $\zeta^{\mu}$ in (5.28). 

The supersymmetry charge algebra is then written as   
\begin{eqnarray}
i\{\tilde{Q}^{\alpha},\tilde{Q}^{\beta}\}
&=&-2(\gamma^{\mu})^{\alpha\beta}P_{\mu}
+\int dx\frac{\hbar g}{2\pi}F(y)
(\gamma^{0})^{\alpha\beta}
\nonumber\\
&&-2Z(\gamma_{5})^{\alpha\beta}
+\int dx \frac{\hbar g}{\pi}\partial_{1}\varphi(y)(\gamma_{5})
^{\alpha\beta}
\end{eqnarray}
where we recall 
\begin{eqnarray}
&&P_{\mu}=\int dx T_{0\mu}(x), 
\ \ \ H=P_{0} ,\nonumber\\
&&Z=\int dx \zeta_{0}(x)
=-\int dx\zeta^{0}(x).
\end{eqnarray}

We thus find that the naive supersymmetry algebra is deformed 
by the effects of the anomaly, and the net effects are the 
replacement 
\begin{eqnarray}
&&H\rightarrow \tilde{H}=H- \int dx\frac{\hbar g}{4\pi}F(x)
,\nonumber\\
&&Z\rightarrow\tilde{Z}=Z- \int dx \frac{\hbar g}{2\pi}
\partial_{1}\varphi(x).
\end{eqnarray}
The modification of 
$H$ is {\em half} of the trace anomaly because 
$\eta_{\mu\nu}F$ 
contributes half as much to $\tilde{T}_{00}$ as to 
$\tilde{T}_{\mu}^{\mu}$. On the other hand, the modification to
$Z$ is the full central charge anomaly, so, as we have argued 
before, $Z$ contains no effects of anomalies.  
This shows that $H$ itself still contains the effects of the 
anomaly.  

By our construction, the above expression gives the same result 
as before for the ground state in the presence of a 
time-independent kink solution.
For example, 
\begin{eqnarray}
\langle \tilde{H}\rangle&=&-\langle\int dx T^{\ 0}_{0}(x)
+\int dx \frac{\hbar g}{4\pi}F(x)\rangle\nonumber\\
&=&-\langle\int dx \tilde{T}^{\ 0}_{0}(x) \rangle
\end{eqnarray}
in terms of the conserved tensor $\tilde{T}^{\ \mu}_{\nu}(x)$
by noting (5.28).
 Similarly  
\begin{eqnarray}
\langle \tilde{Z}\rangle&=&-\langle \int dx\zeta^{0}(x)\rangle
-\langle\int dx \frac{\hbar g}{2\pi}
\partial_{1}\varphi(x) \rangle\nonumber\\
&=&-\langle \int dx\tilde{\zeta}^{0}(x)\rangle
\end{eqnarray}
in terms of the conserved current $\tilde{\zeta}^{\mu}$.

When one solves the classical kink solution, which satisfies 
the BPS bound, one does not distinguish between the operators 
$\tilde{T}_{\mu}^{\ \nu}$ and $\tilde{\zeta}^{\mu}$ or
$T_{\mu}^{\ \nu}$ and $\zeta^{\mu}$. When fully quantized
as in our formulation, these two sets of operators give rise to 
different forms of the supersymmetry algebra. This difference is 
a quantum effect which is not taken account by the 
semi-classical Dirac bracket analysis. Properly distinguishing these 
two sets of currents clarifies the role of 
superconformal anomalies in the saturation of the BPS bound. 
   
\section{Conclusions}
We have presented a superspace description of the anomalies in 
the energy and central charge of a supersymmetric kink. In this 
superspace description these anomalies  form with other 
anomalies a multiplet, and this allows an analysis which 
preserves at all steps rigid ordinary supersymmetry. The 
anomalies occur in Ward identities for various currents, but do 
not always correspond to symmetries of the action. 
The conformal symmetries (Weyl transformations and conformal 
supersymmetry transformations) leave only the free part of the 
action invariant, so that the Ward identities corresponding to 
these transformations are explicitly broken Ward identities. We 
derived them in section 4 by making an ordinary supersymmetry 
transformation in superspace with a parameter which was local 
both in $x$ and in $\theta$. We derived a lemma
(section 3) for the Jacobian of an arbitrary transformation of 
the scalar superfield $\phi$ (not necessary corresponding to a 
symmetry of the action) and these Jacobians produced the 
anomalies in the explicitly broken Ward identities. The 
variation of the action produced the currents in these Ward 
identities, and it turned out to be useful to distinguish between
currents $\tilde{K}^{\mu}=\{\tilde{J}^{\mu}, \tilde{T}_{\mu\nu}, 
\tilde{\zeta}_{\mu}\}$ from the conformal invariant kinetic part
of the action, and currents  $K^{\mu}=\{j^{\mu}, T_{\mu\nu}, 
{\zeta}_{\mu}\}$ from the nonconformal interacting part of the 
action. 
The former depend also on the interactions through the field 
equations of the Heisenberg operators.
The currents $\tilde{K}^{\mu}$ are conserved at the 
quantum level, and these are thus the currents which yield 
time-independent charges of the supersymmetry algebra.
The anomalies involved traces and divergences of the currents
$\tilde{K}^{\mu}$ and $K^{\mu}$. Denoting these contractions 
by $O\tilde{K}$ and $OK$, the anomalies were of the form 
$O\tilde{K}=OK+F+\hbar A_n$, where  $O$ is an (algebraic or  
differential) operator, $F$ a term proportional to the 
fermion field equation and $A_n$ the anomaly.
Since the $OK$ had lower field dimension than the $O\tilde{K}$, 
only the $O\tilde{K}$ contained anomalies. Thus we obtained a 
clear separation in the Ward identities between explicit 
symmetry breaking terms (due to $OK$) and anomalies. 

The general local supersymmetry 
variation in superspace
\begin{equation}
\delta\phi(x,\theta)=\bar{\Omega}(x,\theta)Q\phi(x,\theta)
\end{equation}
generates all the anomalous broken Ward identities.
It contains a term 
$\delta F=-\epsilon^{\mu\nu}c_{\mu}\partial_{\nu}\varphi$
which generates the central charge current and its anomaly,
together with the variation of fermion variable 
$\delta\psi=c_{\mu}\gamma^{\mu}\gamma_{5}\psi$ which is 
anomaly-free. It was then shown that the transformation
\begin{equation}
\delta\phi(x,\theta)=-\delta(\theta)
v^{\mu}(x)\epsilon_{\mu\nu}\partial^{\nu}\phi(x,\theta)
\end{equation}
contains only the  term 
$\delta F=-\epsilon^{\mu\nu}v_{\mu}\partial_{\nu}\varphi$
which generates the central charge current and its anomaly. 
If one uses 
$v^{\mu}=\partial^{\mu}v$ in this variation, one obtains the 
divergence of the central charge current but this yields no
information because a topological current is identically 
conserved. In analogy with 
$U(1)$ gauge theory, the above variation (7.2) corresponds to
the variation
\begin{equation}
A_{\mu}\rightarrow A_{\mu}+v_{\mu}
\end{equation}
to generate the current instead of the variation 
$A_{\mu}\rightarrow A_{\mu}+\partial_{\mu}v$ which generates
the ordinary Noether current. 

Another characterisitic feature of the present formulation is 
that the conserved supersymmetry current 
\begin{equation}
\tilde{J}^{\mu}(x)\equiv j^{\mu}(x)
-\frac{\hbar g}{2\pi}\gamma^{\mu}\psi(x)
\end{equation}
appears in all the local Ward identities for rigid supersymmetry.
Here $j^{\mu}$ stands for the Noether current coming from the 
action and $-\frac{\hbar g}{2\pi}\gamma^{\mu}\psi(x)$ from the 
Jacobian. The
current $j^{\mu}$  is not conserved but free of the 
superconformal gamma-trace anomaly, whereas the conserved current
$\tilde{J}^{\mu}$ contains the gamma-trace anomaly 
$(\gamma_{\mu}\tilde{J}^{\mu})_{anomaly}
=-\frac{\hbar g}{2\pi}\gamma^{\mu}\psi$.
This appearance of two different currents with clear 
anomaly properties makes the analysis of the BPS bound
transparent.

We then obtained the
supersymmetry algebra for the conserved charge 
$\tilde{Q}=\int dx \tilde{J}^{0}(x)$ starting with a 
supersymmetry Ward 
identity by using the BJL prescription.
The algebra is generally modified by the effects of the 
trace and 
central charge anomalies. The deformation of the algebra is 
in such a way that the BPS bound remains saturated
as a result of uniform shifts in energy and central charge in 
the presence of a kink solution.   
If one uses the conserved quantities $\tilde{T}_{\mu}^{\ \nu}$
and $\tilde{\zeta}^{\mu}$, the supersymmetry algebra retains 
the naive form  it had before one incorporates the effects of 
anomalies. If one uses the non-conserved quantities
$T_{\mu}^{\ \nu}$ and $\zeta^{\mu}$, these currents are
 deformed by anomalies, but the physical content remains the 
same. Note that the ``topology'' of the kink vacuum is not 
modified by the superconformal anomalies.
An interesting (and subtle) aspect of the present problem is 
that both the explicit and the anomalous breakings of 
superconformal symmetry are proportional to the coupling 
constant $g$.   

There are only one-loop anomalies  in the present 
(super-renormalizable) theory~\cite{Nastase:1998sy}
\cite{Shifman:1998zy}. 
Thus our analysis of the  
supersymmetry algebra and the BPS bound is exact.   
\\

\noindent {\bf Acknowledgments}
\\

We thank K. Shizuya for helpful discussions of the superspace
evaluation of the Jacobian. One of us (PvN) thanks the particle 
theory group at the University of Tokyo, where the present work 
was initiated, for its 
hospitality. 
The other of us (KF) thanks all the 
members of the C.N.Yang Institute for Theoretical Physics at 
Stony Brook, where this work was completed,  for their 
hospitality.

\appendix

\section{The Bjorken-Johnson-Low method}

We summarize the essence of the BJL method~\cite{bjorken}. 
We start with the correlation function in the path integral 
approach
\begin{equation}
\langle T^{\star}A(x)B(y)\rangle
\end{equation}
where the $T^{\star}$ product is defined for all the space-time
points except for $x^{0}=y^{0}$. In the path integral approach 
one has 
\begin{equation}
\partial_{\mu}\langle T^{\star}A(x)B(y)\rangle=
\langle T^{\star}\partial_{\mu}A(x)B(y)\rangle.
\end{equation}
 Of course this does not hold in general for the canonical 
$T$ product.
 
The basic observation of the BJL prescription is that 
the Fourier transform 
\begin{equation}
\int dx e^{ik(x-y)}\langle T^{\star}A(x)B(y)\rangle
\end{equation}
differs from
\begin{equation}
\int dx e^{ik(x-y)}\langle TA(x)B(y)\rangle
\end{equation}
at most by polynomials in $k_{0}$. We thus fix this
difference by the following condition: If 
\begin{equation}
\lim_{k_{0}\rightarrow\infty}\int dx e^{ik(x-y)}
\langle T^{\star}A(x)B(y)\rangle=0
\end{equation} 
then one can set
\begin{equation}
\int dx e^{ik(x-y)}\langle TA(x)B(y)\rangle
=\int dx e^{ik(x-y)}\langle T^{\star}A(x)B(y)\rangle.
\end{equation} 
On the other hand, if 
(A.5) does not vanish, one defines (A.4) by subtracting 
(A.5)
\begin{eqnarray}
\int dx e^{ik(x-y)}\langle TA(x)B(y)\rangle
&=&\int dx e^{ik(x-y)}\langle T^{\star}A(x)B(y)\rangle
\nonumber\\
&&-\lim_{k_{0}\rightarrow\infty}\int dx e^{ik(x-y)}
\langle T^{\star}A(x)B(y)\rangle
\end{eqnarray}
By construction one has then always  the condition
\begin{equation}
\lim_{k_{0}\rightarrow\infty}\int dx e^{ik(x-y)}
\langle T A(x)B(y)\rangle=0
\end{equation} 
which is a basic property of the $T$ product. The physical 
picture behind this construction is that the expectation value
with the T-product is a smooth function of $x-y$ near 
$x^{0}\sim y^{0}$. Then the Fourier transform is also smooth, 
and tends to zero for large $k_{0}$.

We illustrate this procedure by applying it to the quantization
of the free massive Wess-Zumino model in $d=2$
\begin{eqnarray}
{\cal L}(x)&=& \frac{1}{2}[FF
-\partial^{\mu}\varphi\partial_{\mu}\varphi
-\bar{\psi}\gamma^{\mu}\partial_{\mu}\psi + 2mF\varphi
-m\bar{\psi}\psi].
\end{eqnarray}
The path integral gives the correlation functions
\begin{eqnarray}
&&\langle T^{\star}\varphi(x)\varphi(y)\rangle
=\frac{i}{\partial_{\mu}\partial^{\mu}-m^{2}}\delta(x-y),
\nonumber\\
&&\langle T^{\star}\psi(x)\bar{\psi}(y)\rangle
=\frac{-i}{\delslash+m}\delta(x-y),
\nonumber\\
&&\langle T^{\star}\varphi(x)F(y)\rangle=
\frac{-im}{\partial_{\mu}\partial^{\mu}-m^{2}}\delta(x-y),
\nonumber\\
&&\langle T^{\star}F(x)F(y)\rangle
=i[1+\frac{m^{2}}{\partial_{\mu}\partial^{\mu}-m^{2}}]
\delta(x-y)
\end{eqnarray}
Since (A.5) is satisfied for the $\varphi\varphi$ propagator,
we have 
\begin{equation}
\int dx e^{ik(x-y)}\langle T \varphi(x)\varphi(y)\rangle
=\frac{-i}{k^{2}+m^{2}-i\epsilon}
\end{equation}

{}From this result we can derive the equal time canonical 
commutation relations. First of all, consider the commutator
$[\varphi(x), \varphi(y)]=0$ at equal time. One has, using 
(A.11),
\begin{eqnarray}
0&=&\lim_{k_{0}\rightarrow\infty}k_{0}\int dx 
e^{ik(x-y)}\langle T \varphi(x)\varphi(y)\rangle\nonumber\\
&&=\lim_{k_{0}\rightarrow\infty}(i)\int dx 
e^{ik(x-y)}\frac{\partial}{\partial x^{0}}
\langle T \varphi(x)\varphi(y)\rangle\nonumber\\
&&=\lim_{k_{0}\rightarrow\infty}(i)\{\int dx
e^{ik(x-y)}\langle [\varphi(x), \varphi(y)]\rangle\delta(x^{0}
-y^{0})
\nonumber\\
&&+\int dxe^{ik(x-y)}
\langle T \frac{\partial}{\partial x^{0}}\varphi(x)
\varphi(y)\rangle\}\nonumber\\
&&=\lim_{k_{0}\rightarrow\infty}(i)\int dx
e^{ik(x-y)}\langle [\varphi(x), \varphi(y)]\rangle\delta(x^{0}
-y^{0}).
\end{eqnarray}
We used in the last step that the $T$ product vanishes 
 in the limit $k_{0}\rightarrow\infty$. Because the last line is 
$k_{0}$-independent due to $\delta(x^{0}-y^{0})$ we have  
\begin{eqnarray} 
&&\langle [\varphi(x), \varphi(y)]\rangle\delta(x^{0}
-y^{0})=0
\end{eqnarray}
Using this result in (A.12) without taking the limit 
$k_{0}\rightarrow\infty$ yields
\begin{eqnarray}
&&i\int dxe^{ik(x-y)}
\langle T \frac{\partial}{\partial x^{0}}\varphi(x)
\varphi(y)\rangle=\frac{-ik_{0}}{k^{2}+m^{2}-i\epsilon}.
\end{eqnarray}

By multiplying the expression in (A.14) by $k_{0}$ and 
performing the same analysis as before, we obtain 
\begin{eqnarray}
&&-\int dxe^{ik(x-y)}
\langle [\frac{\partial}{\partial x^{0}}\varphi(x),
\varphi(y)]\rangle\delta(x^{0}-y^{0})\nonumber\\
&&=\lim_{k_{0}\rightarrow\infty}
\frac{-ik^{2}_{0}}{k^{2}+m^{2}-i\epsilon}=i
\end{eqnarray}
Since again this result is $k_{0}$-independent, we have a second
identity
\begin{eqnarray}
\int dx e^{ik(x-y)}\langle T\partial^{2}_{0}\varphi(x)\varphi(y)
\rangle =\frac{-i(k_{1}^{2}+m^{2})}{k^{2}+m^{2}-i\epsilon}
\end{eqnarray}
This can be rewritten as follows by bringing the right-hand
side to the left-hand side 
\begin{equation}
\int dx e^{ik(x-y)}\langle T
(\partial_{\mu}\partial^{\mu}+m^{2})\varphi(x)\varphi(y)\rangle
=0
\end{equation}
which is consistent with the field equation 
$(\partial_{\mu}\partial^{\mu}+m^{2})\varphi(x)=0$. 

Summarizing, we have proven the following commutation relations
\begin{eqnarray}
&&[\frac{\partial}{\partial x^{0}}\varphi(x),
\varphi(y)]\delta(x^{0}-y^{0})=-i\delta^{2}(x-y),
\nonumber\\
&&[\varphi(x),
\varphi(y)]\delta(x^{0}-y^{0})=0.
\end{eqnarray}

The analysis of $\langle T^{\star}\varphi(x)F(y)\rangle$
is essentially the same and we obtain
\begin{eqnarray}
&&[\frac{\partial}{\partial x^{0}}\varphi(x),
F(y)]\delta(x^{0}-y^{0})=im\delta^{2}(x-y),
\nonumber\\
&&[\varphi(x),
F(y)]\delta(x^{0}-y^{0})=0.
\end{eqnarray}
These relations are consistent with the field equation
$F(x)+m\varphi(x)=0$.

As for $\langle T^{\star}F(x)F(y)\rangle$, we have
\begin{eqnarray}
\int dx e^{ik(x-y)}\langle T^{\star}F(x)F(y)\rangle
=i[1- \frac{m^{2}}{k^{2}+m^{2}-i\epsilon}]
\end{eqnarray}
and thus the BJL method gives according to (A.7)
\begin{eqnarray}
\int dx e^{ik(x-y)}\langle TF(x)F(y)\rangle
=-i\frac{m^{2}}{k^{2}+m^{2}-i\epsilon}
\end{eqnarray}
Using the same steps as before, one finds 
\begin{eqnarray}
&&[\frac{\partial}{\partial x^{0}}F(x),
F(y)]\delta(x^{0}-y^{0})=-im^{2}\delta^{2}(x-y),
\nonumber\\
&&[F(x),
F(y)]\delta(x^{0}-y^{0})=0.
\end{eqnarray}
Also, these quantization conditions are consistent with 
the field equation $F=-m\varphi(x)$.

We now come to the fermionic sector
\begin{eqnarray}
\int dxe^{ik(x-y)}\langle T^{\star}\psi(x)\bar{\psi}(y)\rangle
=\frac{-i}{-i\kslash+m}
\end{eqnarray}
and thus 
\begin{eqnarray}
\int dxe^{ik(x-y)}\langle T\psi(x)\bar{\psi}(y)\rangle
=\frac{-i}{-i\kslash+m}=-i\frac{i\kslash+m}{k^{2}+m^{2}}
\end{eqnarray}
Proceeding as before we obtain
\begin{eqnarray}
&&\lim_{k_{0}\rightarrow\infty}k_{0}
\int dxe^{ik(x-y)}\langle T\psi(x)\bar{\psi}(y)\rangle
\nonumber\\
&&=
i\lim_{k_{0}\rightarrow\infty}
\int dxe^{ik(x-y)}\frac{\partial}{\partial x^{0}}
\langle T\psi(x)\bar{\psi}(y)\rangle
\nonumber\\
&&=
i\lim_{k_{0}\rightarrow\infty}
\int dxe^{ik(x-y)}[\langle \{\psi(x),\bar{\psi}(y)\}\rangle
\delta(x^{0}-y^{0})
+\langle T\frac{\partial}{\partial x^{0}}\psi(x)
\bar{\psi}(y)\rangle]
\nonumber\\
&=&\lim_{k_{0}\rightarrow\infty}
-ik_{0}\frac{i\kslash+m}{k^{2}+m^{2}}=-\gamma^{0}
\end{eqnarray}
We thus obtain the quantization condition
\begin{equation}
\{\psi(x),\bar{\psi}(y)\}
\delta(x^{0}-y^{0})=i\gamma^{0}\delta(x-y)
\end{equation}
and
\begin{equation}
\int dxe^{ik(x-y)}\langle T (\delslash+m)\psi(x)\bar{\psi}(y)
\rangle=0
\end{equation}
which is consistent with the field equation 
$(\delslash+m)\psi(x)=0$. Note that (A.26) gives the same result 
as obtained from Dirac quantization, where the conjugate 
momentum of $\psi$ equals $-\frac{i}{2}\psi$, but the Dirac 
bracket removes this factor $1/2$. 

When one applies the BJL prescription to perturbation theory,
one should define the correlation functions in terms of 
unrenormalized (bare) fields. (Recall that the canonical 
formalism is phrased in terms of Heisenberg fields which are
unrenormalized.)
Since the canonical 
equal-time commutator is based on 
the basic assumption that it is not modified by interactions,
one should also use bare fields in perturbation theory, with a
momentum cut-off in the loop, in order that one reproduces by 
the BJL 
approach the results of the canonical formalism. 
However, one can incorporate the effects of anomalies 
naturally into the BJL analysis.

We illustrate the BJL analysis in the presence of anomalies,
which goes beyond the naive canonical formulation, by taking 
the identity (5.13) as an example (although the anomaly in this 
case does not have a physical significance for rigid 
supersymmetry). We started  in (5.13) from 
\begin{equation}
-i\partial_{\mu}\langle T^{\star}J^{\mu,\alpha}(x)
\phi(y,\theta)
\rangle
+\langle \delta_{susy}\phi(y,\theta)\rangle=0
\end{equation}
where
\begin{equation}
J^{\mu}=j^{\mu}- \frac{\hbar g}{2\pi}\gamma^{\mu}\psi(x)
\end{equation}
and $\delta_{susy}\phi(y,\theta)=\delta^{2}(x-y)
Q^{\alpha}\phi(x,\theta)$.
The BJL subtraction procedure leads to the result that 
 the derivative is outside the T-product: One writes down all
 Feynman graphs with $j^{\mu}$ and 
$- \frac{\hbar g}{2\pi}\gamma^{\mu}\psi(x)$, uses a cut-off
$\Lambda$ in each loop and takes the limit 
$k_{0}\rightarrow\infty$. The result is   
\begin{equation}
-i\partial_{\mu}\langle T J^{\mu,\alpha}(x)\phi(y,\theta)\rangle
+\langle \delta^{2}(x-y)Q^{\alpha}\phi(x,\theta)\rangle=0.
\end{equation}
This implies
\begin{equation}
i[J^{0,\alpha}(x),\phi(y,\theta)]\delta(x^{0}-y^{0})
=\delta^{2}(x-y)[\frac{\partial}{\partial\bar{\theta}_{\alpha}}
-(\gamma^{\mu}\theta)^{\alpha}\partial_{\mu}]\phi(x,\theta)
\end{equation}
and $\langle T\partial_{\mu}J^{\mu,\alpha}(x)\phi(y,\theta)
\rangle=0$ which in turn implies the current conservation
$\partial_{\mu}J^{\mu,\alpha}(x)=0$. 

If one combines this commutation relation with the anomaly
relation (A.29), one obtains the anomalous commutation relation
for the current $j^{\mu}$
\begin{eqnarray}
i[j^{0,\alpha}(x),\phi(y,\theta)]\delta(x^{0}-y^{0})
&=&i[\frac{\hbar g}{2\pi}(\gamma^{0}\psi(x))^{\alpha},
\phi(y,\theta)]
\delta(x^{0}-y^{0})\nonumber\\
&&+\delta^{2}(x-y)[\frac{\partial}{\partial\bar{\theta}_{\alpha}}
-(\gamma^{\mu}\theta)^{\alpha}\partial_{\mu}]\phi(x,\theta).
\end{eqnarray}
The term with $\hbar g$ is the operator counterpart of the 
non-trivial Jacobian in the path integral formulation.   

A special property of the anomaly (unlike the induced effect
such as the anomalous magnetic moment in QED) is that we have a 
well-defined (local) operator relation expressed by (A.29), 
which is 
established in some cases up to all orders in perturbation 
theory, and that the anomaly itself is finite and independent of 
the cut-off parameter. In the above BJL analysis of (A.32),
we regarded both of (A.29) and (A.31) as exact (bare) operator 
relations (though we worked only to the one-loop accuracy in 
the present paper).

\end{document}